# Assessment of Ammonia as a Biosignature Gas in Exoplanet Atmospheres

Jingcheng Huang[1,2], Sara Seager[2,3,4], Janusz J. Petkowski[2], Sukrit Ranjan[2], Zhuchang Zhan[2]

[1]Department of Chemistry, Massachusetts Institute of Technology
[2]Department of Earth, Planetary and Atmospheric Sciences, Massachusetts Institute of Technology
[3]Department of Physics, Massachusetts Institute of Technology
[4]Department of Aeronautics and Astronautics, Massachusetts Institute of Technology

## Abstract

Ammonia ($NH_3$) in a terrestrial planet atmosphere is generally a good biosignature gas, primarily because terrestrial planets have no significant known abiotic $NH_3$ source. The conditions required for $NH_3$ to accumulate in the atmosphere are, however, stringent. $NH_3$'s high water solubility and high bio-useability likely prevent $NH_3$ from accumulating in the atmosphere to detectable levels unless life is a net source of $NH_3$ and produces enough $NH_3$ to saturate the surface sinks. Only then can $NH_3$ accumulate in the atmosphere with a reasonable surface production flux.

For the highly favorable planetary scenario of terrestrial planets with $H_2$-dominated atmospheres orbiting M dwarf stars (M5V), we find a minimum of about 5 ppm column-averaged mixing ratio is needed for $NH_3$ to be detectable with JWST, considering a 10 ppm JWST systematic noise floor. When the surface is saturated with $NH_3$ (i.e., there are no $NH_3$-removal reactions on the surface), the required biological surface flux to reach 5 ppm is on the order of $10^{10}$ molecules $cm^{-2}$ $s^{-1}$, comparable to the terrestrial biological production of $CH_4$. However, when the surface is unsaturated with $NH_3$, due to additional sinks present on the surface, life would have to produce $NH_3$ at surface flux levels on the order of $10^{15}$ molecules $cm^{-2}$ $s^{-1}$ (~$4.5\times10^6$ Tg $year^{-1}$). This value is roughly 20,000 times greater than the biological production of $NH_3$ on Earth and about 10,000 times greater than Earth’s $CH_4$ biological production.

Volatile amines have similar solubilities and reactivities to $NH_3$ and hence share $NH_3$'s weaknesses and strengths as a biosignature. Finally, to establish $NH_3$ as a biosignature gas, we must rule out mini-Neptunes with deep atmospheres, where temperatures and pressures are high enough for $NH_3$’s atmospheric production.

## 1. Introduction

A growing part of exoplanet research focuses on the search for biosignature gases with future telescopes. Biosignature gases are gases produced by living organisms either as final products or by-products from biochemical pathways. A viable biosignature gas must have spectral features that are both detectable and distinguishable from other molecular species (e.g., Zhan et al., 2021). A growing number of gases have been proposed as biosignatures. The most well-known example is $O_2$ (e.g., Jeans, 1930; Meadows et al., 2018). Other biosignature gas candidates include DMS ($(CH_3)_2S$) (Domagal-Goldman et al., 2011; Seager et al., 2012; Arney et al., 2018), methane ($CH_4$) (Leger et al., 1996; Des Marais et al., 2002; Kaltenegger et al., 2007; Dlugokencky et al., 2011), nitrous oxide ($N_2O$) (Des Marais et al., 2002; Tian et al., 2015), ammonia ($NH_3$) (Seager et al., 2013), methyl chloride ($CH_3Cl$) (Segura et al., 2005), phosphine ($PH_3$) (Sousa-Silva et al., 2019), and isoprene ($C_5H_8$) (Zhan et al., 2021).

We are motivated to study the biosignature potential of $NH_3$ since $NH_3$ plays a significant role in biochemistry. Due to its high bio-usability, plants and various microorganisms can easily absorb $NH_3$. $NH_3$ is an ideal nitrogen source for life on Earth since it can be integrated into various amino acids and other organic molecules without life having to break the strong $N_2$ triple bond. Additionally, some life can use $NH_3$ as an energy source by oxidizing $NH_3$. Furthermore, $NH_3$ stands out from all the previously studied biosignature gases because $NH_3$ has a very high solubility in water. $NH_3$'s high water solubility means the atmospheric accumulation of $NH_3$ is highly dependent on planetary conditions such as whether life produces a substantial amount of $NH_3$ and whether there are active $NH_3$-removal sinks on the surface (including land and ocean). If life does not produce enough $NH_3$ to saturate the surface sinks, rain can much more efficiently remove $NH_3$ from the atmosphere than other atmospheric gases.

$NH_3$ has been studied before. Early studies suggest $NH_3$ is unlikely to be present in an early Earth atmosphere because it is easily destroyed photochemically (Kasting et al., 1982; Sagan and Chyba, 1997). In a separate paper, we studied $NH_3$ as a biosignature gas for a specific, optimistic planetary scenario of a planet with an $H_2$ - $N_2$ atmosphere termed the 'cold Haber World' (Seager et al., 2013). On such a planet, life may have evolved catalytic machinery to break the $N_2$ triple bond and extract energy by fixing atmospheric $H_2$ and $N_2$ into $NH_3$ (Seager et al., 2013). A somewhat similar catalytic process occurs in the industry (the Haber process) at specific temperatures and pressures.

In this work, we reassess $NH_3$ as a biosignature gas by considering different planetary scenarios and surface conditions. We first summarize $NH_3$'s sources and sinks on Earth in Section 2. We then describe our models in Section 3. We present our results in Section 4. Finally, we conclude our paper with a discussion of our results and a description of limitations in Section 5.

## 2. Ammonia Sources and Sinks

To assess the biosignature potential of $NH_3$, we first need to review how $NH_3$ is produced (Section 2.1) and destroyed (Section 2.2) on Earth. $NH_3$ sources on Earth are few and either anthropogenic (Section 2.1.1) or biological (Section 2.1.2), with limited abiotic sources (Section 2.1.3). In contrast, $NH_3$ has many destruction pathways. We discuss $NH_3$ atmospheric sinks in Section 2.2.1 and $NH_3$ biological sinks in Section 2.2.2.

### 2.1 Ammonia Sources on Earth

Ammonia is a trace gas (< 1 ppbv) in modern Earth's atmosphere, with a residence time of less than a day (Zhu et al., 2015). Current data on $NH_3$ emissions at a global level have significant uncertainties due to the lack of measurements in most parts of Earth's atmosphere (Olivier et al., 1998; Bouwman et al., 2002; Zhu et al., 2015; Van Damme et al., 2018). It is worth noting that anthropogenic $NH_3$ production is a significant $NH_3$ source on Earth (Söderlund et al., 1976; Dentener and Crutzen, 1994; Olivier et al., 1998).

### 2.1.1 Anthropogenic $NH_3$ Production

On Earth, most global $NH_3$ emissions are due to anthropogenic activities, especially agriculture (Behera et al., 2013). Large quantities of ammonia are directly applied on farm fields as fertilizer (anhydrous $NH_3$) or are used to produce secondary nitrogen fertilizers. Besides anhydrous $NH_3$, synthetic N fertilizers such as urea ($(NH_2)_2CO$) and ammonium bicarbonate ($NH_4HCO_3$) are also popular N fertilizers (Behera et al., 2013). Both applications result in very high localized $NH_3$ emissions from farmlands (e.g., Dong et al., 2009; Hou and Yu, 2020). $NH_3$ emissions from N fertilizers depend on many factors, including the type of fertilizers, local meteorological conditions, and soil properties (Behera et al., 2013). It has been estimated that up to 10% of N content is lost through $NH_3$ emission when anhydrous $NH_3$ fertilizer is used. This number rises to 18.7% for urea fertilizer (Behera et al., 2013). Animal husbandry also significantly contributes to global $NH_3$ emissions (through the microbial decomposition of animal waste, see section 2.1.2 below). Coal-burning also produces $NH_3$, even though the quantity is relatively small (Soderlund et al., 1976, Behera et al., 2013).

Industrial $NH_3$ sources come from the production of nitric acid, fuel, explosives, and refrigerants. Other minor global $NH_3$ emissions include agriculture-related biomass decomposition and motor vehicles (Zhu et al., 2015). We summarize the major $NH_3$ sources and their flux in Table 2-1.

**Table 2-1.** Major anthropogenic $NH_3$ sources and their estimated flux (Soderlund et al., 1976).

| Ammonia sources | Rate [Tg/year] |
|---|---|
| Domestic animals and human | 20 - 35 |
| Burning of coal | 4 - 12 |
| Wild animals | 2 - 6 |
| Total | 26 - 53 |

Global $NH_3$ emission flux is extremely hard to estimate because most $NH_3$ measurements are local, and $NH_3$ abundance varies widely (Olivier et al., 1998; Zhu et al., 2015; Van Damme et al., 2018). It has been reported that global $NH_3$ emissions may have quadrupled in the past few years due to the increasing demand for food and other agricultural products (Lamarque et al., 2011 and Zhu et al., 2015). Global $NH_3$ emissions will likely keep growing in the near future (Moss et al., 2010, Behera et al., 2013). The fact that the $NH_3$ emission rate has a strong dependency on climate (Sutton et al., 2013) makes an accurate $NH_3$ emission estimate even more difficult. We summarize some estimates of global anthropogenic $NH_3$ emission from the past few decades in Table 2-2.

**Table 2-2.** A summary of global anthropogenic $NH_3$ emission estimates (Zhu et al., 2015).

| Global $NH_3$ emission rate [Tg/year] | Source |
|---|---|
| 38.5 | Moss et al., 2010 |
| 53 | Soderlund et al., 1976 |
| 54 | Bouwman et al., 1997 |

As an aside, anthropogenic $NH_3$ is produced by the Haber-Bosch process. The Haber-Bosch process reduces unreactive atmospheric nitrogen ($N_2$) to ammonia with molecular hydrogen ($H_2$). This reaction requires a metal catalyst (e.g., iron and other catalysts), high temperatures (450 ~ 550 °C), and high pressures (250 ~ 350 bar).

$$3H_2 + N_2 \rightarrow 2NH_3 \quad \Delta G_f^o = -32.8 \text{ kJ/mol} \qquad (2\text{-}1)$$

The Haber-Bosch reaction has a standard Gibbs free energy change ($\Delta G_f^o$) of -32.8 kJ/mol and standard enthalpy change ($\Delta H_f^o$) of -91.8 kJ/mol. Even though this reaction is thermodynamically favorable at room temperatures (~300 K), it is kinetically disfavored. The use of catalysts significantly lowers the activation energy required for breaking the triple bond in the $N_2$ gas. It is worth noting that the reaction becomes thermodynamically unfavorable at high temperatures (~550°C). Nevertheless, with high pressures, the forward reaction proceeds fast enough to give an industrially acceptable $NH_3$ yield.

### 2.1.2 Biological $NH_3$ Production

Three known biological processes generate ammonia (Figure 1). We briefly summarize them below.

Nitrogen fixation: Biological nitrogen fixation by diazotrophic bacteria and Archaea converts atmospheric $N_2$ to $NH_3$. The chemical reaction is catalyzed by a nitrogenase enzyme's metal cluster iron-molybdenum cofactor (FeMoco) (Burgess et al., 1996). The biosynthesis of $NH_3$ is highly energetically costly and is coupled to the hydrolysis of 16 ATP molecules.

$$N_2 + 8H^+ + 8e^- + 16\text{Mg-ATP} \rightarrow 2NH_3 + H_2 + 16\text{Mg-ADP} + 16P_i \qquad (2\text{-}2)$$

Note that because the nitrogenase enzyme is susceptible to $O_2$, many nitrogen-fixing organisms inhabit strictly anaerobic conditions. Despite its complex catalytic mechanism, biological nitrogen fixation likely originated very early in the evolution of life on Earth. Nitrogen isotopic studies suggest that molybdenum-based nitrogenase arose earlier than 3.2 Gya (Stüeken et al., 2015).

Ammonification: Many microorganisms convert organic nitrogen within organic matter (e.g., animal and plant waste, etc.) into $NH_3$. Bacterial and fungal ammonification of animal waste is one of the main processes contributing to the agricultural flux of atmospheric $NH_3$ (see section 2.1.1).

Dissimilatory nitrate reduction to ammonium (DNRA): Dissimilatory nitrate reduction to ammonium (DNRA; nitrate/nitrite ammonification) is a process of anaerobic chemoorganoheterotrophic respiration that uses nitrate ($NO_3^-$), instead of $O_2$, as a final electron acceptor (Lam et al., 2011 and Kraft et al., 2011). As a result of DNRA, organic matter is oxidized anaerobically with a simultaneous reduction of nitrate to nitrite and subsequently to ammonia (or ammonium ion) ($NO_3^- \rightarrow NO_2^- \rightarrow NH_4^+$) (Lam et al., 2011). Dissimilatory nitrate reduction to ammonium is typically found in prokaryotes. However, eukaryotic microorganisms can also carry out DNRA (Kuypers et al., 2018; Preisler et al., 2007; and Stief et al., 2014). Note that in contrast to denitrification (the production of $N_2$ gas), the DNRA process conserves nitrogen, as soluble $NH_4^+$ can be efficiently utilized in biochemical processes (Marchant et al., 2014).

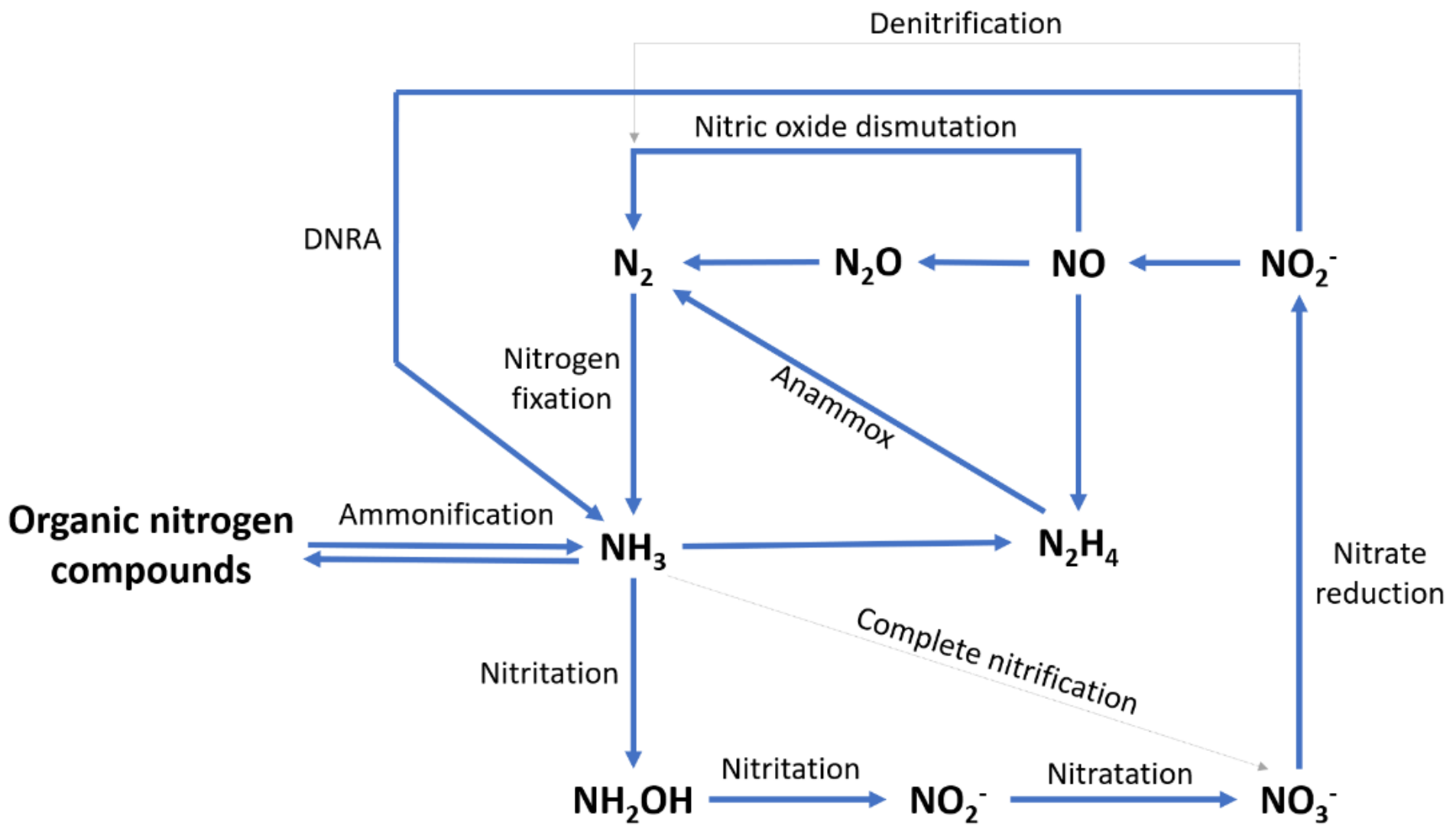

**Figure 1.** Interconversions between the main nitrogen species.

Of the three processes above (excluding ammonification that results from human industrialized agricultural activity (see section 2.1.1)), biological nitrogen fixation at 200 Tg/year dominates the biological production of $NH_3$ (Rascio and Rocca 2013). We compare the annual biological production of ammonia ($NH_3$) to that of nitrogen gas ($N_2$), methane ($CH_4$), isoprene ($C_5H_8$), nitrous oxide ($N_2O$), and methyl chloride ($CH_3Cl$) in Figure 2.

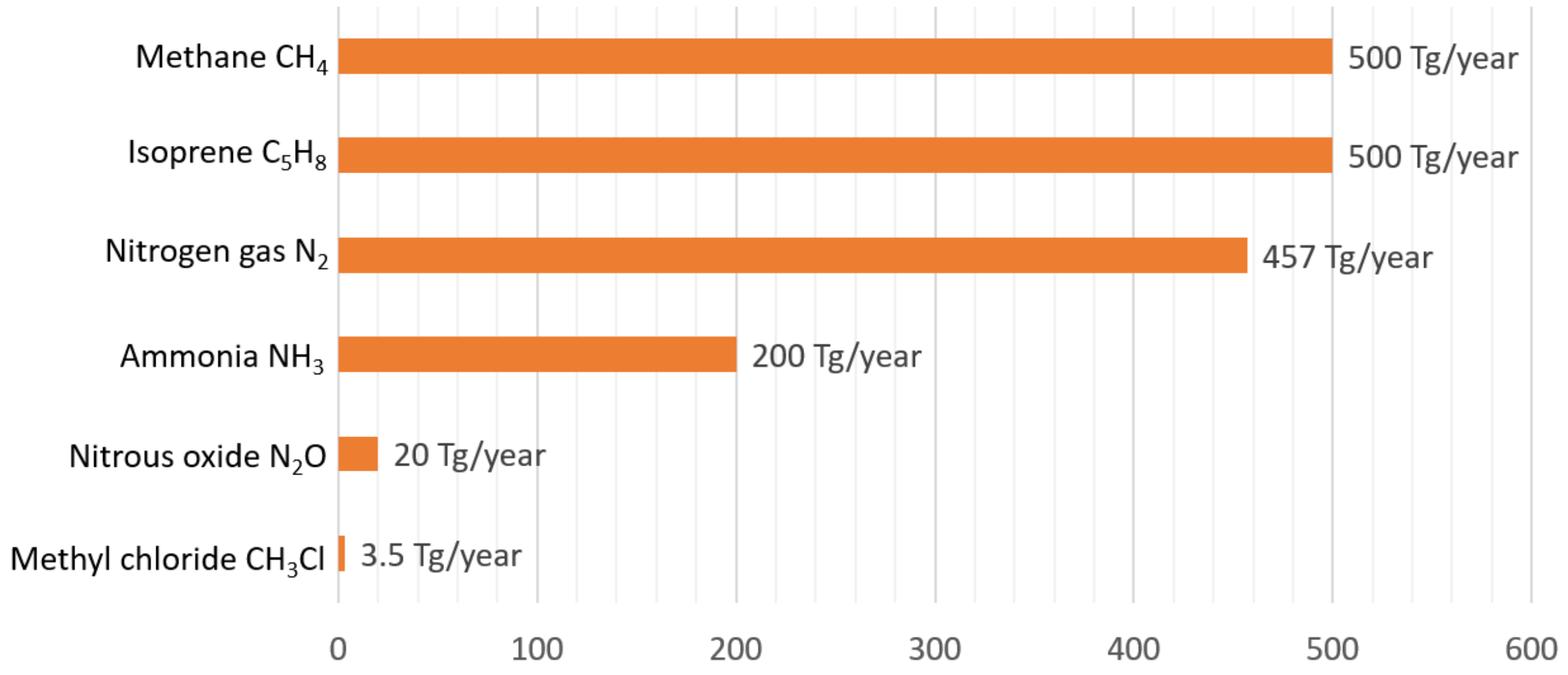

**Figure 2.** Biological production of six different gases in Tg year$^{-1}$. The biological production of ammonia is significantly smaller than that of methane or isoprene. Specifically, the bioproduction of ammonia is only about half of that of biologically produced nitrogen gas. References: Isoprene (Zhan et al., 2021); methane (Guenther et al., 2006); nitrogen gas (Yeung et al., 2019); ammonia (Rascio and Rocca, 2013); nitrous oxide and methyl chloride: (Tian et al., 2015), (Yokouchi et al., 2015) and (Korhonen et al., 2008).

### 2.1.3 Minor $NH_3$ Abiotic Sources

There are very few minor abiotic $NH_3$ sources on Earth[1]. We briefly summarize them below.

$NH_3$ is a trace component of volcanic gases erupted by some, but not all, volcanoes on Earth (see (Möller 2014) and their Table 2.38). For example, (Uematsu et al., 2004) reported an emission of volcanic $NH_3$ from the Miyake-jima volcano that reached 5 ppb locally. The volcanic $NH_3$ is likely emitted due to the thermal breakdown of organic matter of biological origin (Sigurdsson et al., 2015), rather than the reduction of more oxidized N species, like $N_2$. Due to Earth's oxidizing conditions, volcanic outgassing includes some oxidized nitrogen species such as NO and $HNO_3$ (Mather et al., 2004, Martin et al., 2007, Gaillard & Scaillet, 2014). In contrast to Earth, on a planet with a reducing upper mantle condition, nitrogen would be predominately in the form of $NH_4^+$ in the fluids and could be degassed as $NH_3$ (Mikhail et al., 2014 and Mikhail et al., 2017). Detailed modeling is required to explore the plausibility of this scenario further.

$NH_3$ can be photochemically produced on iron-doped $TiO_2$ containing sands without microorganisms' aid (Kasting et al., 1982; Schrauzer et al., 1983). When exposed to sunlight or UV, $TiO_2$ containing sands can reduce $N_2$ to $NH_3$ and trace $N_2H_4$. However, only up to $10^7$ tons (or 10 Tg) of $NH_3$ can be photochemically produced on desert sands every year (Schrauzer et al. 1983). Furthermore, this abiotic $N_2$ photo-fixation is only significant in arid and semiarid regions but not in areas with abundant rainfall and covered by vegetation, where biological $N_2$ fixation dominates (Schrauzer et al. 1983).

---

[1] See Supplementary Information SI.8 for our $NH_3$ abiotic false-positive analysis.

In the presence of dissolved $Ni_3Fe$, NiFe, $Ni^0$, and $Fe^0$ in the ocean, $N_2$, $NO_2^-$ and $NO_3^-$ can be converted to $NH_3/NH_4^+$ (Smirnov et al. 2008). This potential abiotic $NH_3/NH_4^+$ synthesis can happen in anaerobic (oxygen-depleted) conditions or reducing ($H_2$-rich) environments. Such conditions can potentially be present around the Hadean submarine hydrothermal vents (Smirnov et al. 2008). However, this reduction process is strongly temperature-dependent and only becomes efficient at high temperatures (~200°C and above).

$NH_3$ is very unlikely to be produced in hydrothermal vents on modern Earth, and even if it were, $NH_3$ would dissolve in Earth's ocean. The hydrothermal vents are not known to reach conditions for net $NH_3$ synthesis. There may be one extreme exception: the Piccard hydrothermal field. The Piccard hydrothermal system is the deepest known seafloor hot-spring (4957 ~ 4987 m) with pressures (~500 bar), temperatures (~500 °C), and $H_2$ gas abundances (from serpentinization) to enable some limited $N_2$ reduction (McDermott et al., 2018) if a suitable catalyst is present.

Lightning might produce $NH_3$ in the Earth's atmosphere, but only in tiny amounts if at all (see Section 5.3 for further discussion). In Earth's atmosphere, most of $NH_3$ is generated (or 'recycled') from $NH_2$ radicals, which likely originate from the photolysis of $NH_3$ itself (see Table 2-3). Hence, we do not consider Earth's atmospheric chemistry as a net source of $NH_3$. For completeness, we include the high-temperature reactions in Table 2-3 even though the rates are extremely low at Earth temperatures.

**Table 2-3.** $NH_3$-producing reactions. The reaction rates are in units of $cm^3 \cdot molecule^{-1} \cdot s^{-1}$.

| Chemical Reaction | Rate at 298K | Rate law | Source |
|---|---|---|---|
| $H_2 + NH_2 \rightarrow NH_3 + H$ | High temp reaction | $6.75 \cdot 10^{-14} \cdot (T/298)^{2.6} \cdot \exp[-3006.8/T]$ | NIST |
| $NH_2 + C_2H_2 \rightarrow C_2H + NH_3$ | High temp reaction | $8.2 \cdot 10^{-13} \cdot \exp[-2780/T]$ | NIST |
| $NH_2 + C_2H_4 \rightarrow C_2H_3 + NH_3$ | $4.07 \times 10^{-16}$ | $3.42 \cdot 10^{-14} \cdot \exp[-1320.6/T]$ | NIST |
| $NH_2 + C_2H_4O \rightarrow C_2H_3O + NH_3$ | $5.28 \times 10^{-15}$ | $3.50 \cdot 10^{-13} \cdot \exp[-1250/T]$ | NIST |
| $NH_2 + C_2H_6 \rightarrow C_2H_5 + NH_3$ | $2.13 \times 10^{-18}$ | $2.74 \cdot 10^{-14} \cdot (T/298)^{3.46} \cdot \exp[-2820/T]$ | NIST |
| $NH_2 + CH_4 \rightarrow CH_3 + NH_3$ | $6.90 \times 10^{-18}$ | $8.77 \cdot 10^{-15} \cdot (T/298)^{3} \cdot \exp[-2130/T]$ | NIST |
| $NH_2 + H_2O \rightarrow OH + NH_3$ | $9.52 \times 10^{-22}$ | $2.1 \cdot 10^{-13} \cdot (T/298)^{1.9} \cdot \exp[-5725/T]$ | NIST |
| $NH_2 + HO_2 \rightarrow NH_3 + O_2$ | $3.40 \times 10^{-11}$ | $3.4 \cdot 10^{-11}$ | JPL |
| $N_2H_3 + N_2H_3 \rightarrow 2NH_3 + N_2$ | High temp reaction | $5.0 \cdot 10^{-12}$ | NIST |
| $NH_2 + OH \rightarrow NH_3 + O$ | $1.43 \times 10^{-13}$ | $3.32 \cdot 10^{-13} \cdot (T/298)^{0.4} \cdot \exp[-250.2/T]$ | NIST |
| $H + NH_2 \rightarrow NH_3$ | $3.00 \times 10^{-30}$ | $3.00 \cdot 10^{-30}$ | NIST |

## 2.2 Ammonia Sinks on Earth

Compared to $NH_3$'s few formation pathways on Earth, there are numerous routes through which $NH_3$ can be removed. This section focuses on $NH_3$ atmospheric sinks (Section 2.2.1) and biological sinks (Section 2.2.2).

### 2.2.1 $NH_3$ Atmospheric Sinks

$NH_3$ is highly reactive in Earth's atmosphere and is destroyed by direct photolysis and reactions with UV-generated radicals. In Earth's oxic atmosphere, OH is the dominant radical. In exoplanet atmospheres with different bulk compositions, other radicals may become dominant. In particular, for an $H_2$-dominated (highly reducing) atmosphere, the most abundant reactive species are H radicals, produced by photolysis of atmospheric water vapor (Hu et al., 2012). In $N_2$ or $CO_2$-dominated atmospheres, O may also be important, though this needs to be further explored (Hu et al., 2012; Ranjan et al., 2020). We list the reactions between $NH_3$ and O, H, and OH radicals in Table 2-4.

**Table 2-4.** Reactions between $NH_3$ and H, O and OH radicals in the atmosphere. The reaction rates have units of $cm^3 \cdot molecule^{-1} \cdot s^{-1}$. (Reference: NIST)

| Reactions | Rate at 298K | Rate law |
|---|---|---|
| $NH_3 + H\cdot \rightarrow H_2 + NH_2$ | $8.27 \times 10^{-20}$ | $4.27 \cdot 10^{-14} \cdot (T/298)^{3.87} \cdot \exp[-32.59/RT]$ |
| $NH_3 + O\cdot \rightarrow \cdot OH + NH_2$ | $4.37 \times 10^{-17}$ | $2.87 \cdot 10^{-13} \cdot (T/298)^{2.10} \cdot \exp[-21.78/RT]$ |
| $NH_3 + \cdot OH \rightarrow H_2O + NH_2$ | $1.57 \times 10^{-13}$ | $3.50 \cdot 10^{-12} \cdot \exp[-7.69 \pm 1.66/RT]$ |

$NH_3$ can also photodissociate under UV conditions. The photolysis rate coefficient $J_A$ can be quantified by (e.g., Brasseur et al., 2017):

$$J_A = \int q_\lambda \cdot I_\lambda \cdot \sigma_\lambda \cdot e^{-\tau_\lambda}\, d\lambda \qquad (2\text{-}3)$$

Here $q_\lambda$ is the quantum yield of $NH_3$ photolysis. $I_\lambda$ is the solar intensity at the top of the atmosphere, $\tau_\lambda$ is the optical depth, and $\sigma_\lambda$ is the absorption cross-section of $NH_3$. Ammonia photolysis has two pathways, which we summarize in Table 2-5. Reaction rates are computed in (Hu et al., 2012).

**Table 2-5.** Photolysis reactions of $NH_3$ (Hu et al., 2012)

| Reaction | Quantum yields | Reaction rate at 295 K [$s^{-1}$] |
|---|---|---|
| $NH_3 + h\nu \rightarrow NH_2 + H$ | <106nm: 0.3<br>106-165 nm: linear interpolation<br>>165nm: 1.0 | $6.89 \times 10^{-5}$ |
| $NH_3 + h\nu \rightarrow NH + H_2$ | <106nm: 0.7<br>106-165 nm: linear interpolation<br>>165nm: 0 | $1.55 \times 10^{-6}$ |

Atmospheric $NH_3$ will react with nonradical molecules in the atmosphere. In total, we identified five other reactions that can remove $NH_3$ (Table 2-6), and this list is complete for reactions that remove $NH_3$, according to the NIST database (Manion et al., 2008). As in Table 2-3, we include the high-temperature reactions for completeness.

**Table 2-6.** $NH_3$-removing reactions. The reaction rates are in units of $cm^3 \cdot molecule^{-1} \cdot s^{-1}$.

| Chemical Reaction | Rate at 298K | Rate law | Source |
|---|---|---|---|
| $NH + NH_3 \rightarrow NH_2 + NH_2$ | High temp reaction | $5.25 \cdot 10^{-10} \cdot \exp[-13470/T]$ | NIST |
| $NH_3 + CH \rightarrow HCN + H_2 + H$ | $2.50 \times 10^{-11}$ | $7.24 \cdot 10^{-11} \cdot \exp[-317/T]$ | NIST |
| $NH_3 + CH_3 \rightarrow CH_4 + NH_2$ | High temp reaction | $9.55 \cdot 10^{-14} \cdot \exp[-4895/T]$ | NIST |
| $NH_3 + CN \rightarrow HCN + NH_2$ | $1.66 \times 10^{-11}$ | $1.66 \cdot 10^{-11}$ | NIST |
| $NH_3 + CNO \rightarrow CO + N_2H_3$ | $1.20 \times 10^{-14}$ | $6.29 \cdot 10^{-14} \cdot (T/298)^{2.48} \cdot \exp[-493/T]$ | NIST |

### 2.2.2 $NH_3$ Biological Sinks

On Earth, life is a significant sink for $NH_3$. Three biological processes are critical for $NH_3$ depletion by life: nitrification, anammox, and direct assimilation of $NH_3$ into nitrogen-containing organic building blocks of life (i.e., amino acids) (Figure 1). We discuss nitrification and anammox briefly below.

Nitrification is a process of oxidation of ammonia ($NH_3$) to nitrate ($NO_3^-$) (Wendeborn 2020). $NH_3$ oxidizing bacteria and archaea harvest energy by aerobic oxidation of $NH_3$ to $NO_3^-$ via a two-step process (Schimel et al., 2004). In the first step $NH_3$ is oxidized to $NO_2^-$ by chemoautotrophs according to the following formula: $NH_3 + O_2 + 2e^- + 2H^+ \rightarrow NO_2^- + 4e^- + 5H^+$ ($\Delta G° = -275$ kJ mol$^{-1}$) (Daims et al., 2015, Caranto et al., 2017). In the second step $NO_2^-$ is further oxidized to nitrate: $NO_2^- + H_2O \rightarrow NO_3^- + 2H^+ + 2e^-$ ($\Delta G° = -74$ kJ mol$^{-1}$) (Schimel et al., 2004; Daims et al., 2015). Nitrification is performed primarily by soil bacteria (e.g., *Nitrosomonas sp., Nitrospira sp.*).

The second major biological sink of $NH_3$ is called anammox. Anammox is a process of anaerobic oxidation of $NH_3$. Obligately anaerobic chemolithoautotrophs can perform anammox. (Kartal et al., 2011; Van Niftrik et al., 2012). According to the following formula, anammox-capable bacteria convert nitrite and ammonia directly to $N_2$ gas: $NH_4^+ + NO_2^- \rightarrow N_2 + 2H_2O$ ($\Delta G° = -357$ kJ·mol$^{-1}$). Anammox is especially prominent in the oceans, even around hydrothermal vents along the Mid-Atlantic Ridge (Byrne 2009). Anammox is responsible for 50% of marine $N_2$ production (Strous et al., 2006; Devol, 2004).

## 3. Methods

We now describe our methods for assessing $NH_3$ as a biosignature gas. To study the biosignature potential of $NH_3$, we have to determine under which conditions can $NH_3$ accumulate in the atmosphere, considering $NH_3$'s high water solubility and high bio-usability. We first review solubility and Henry's law (Section 3.1). We then introduce our simplified ocean-$NH_3$ interaction model (Section 3.2), followed by a description of our one-dimensional photochemistry model (Section 3.3) and our transmission spectra model (Section 3.4).

### 3.1 Solubility and Henry's law

We use Henry's law to study the solubility of $NH_3$. There are multiple ways to define Henry's law constant (R. Sander 2015). We choose the following form.

$$H_{(X)}^{CP} = \frac{C_{(X)}}{P} \tag{3-1}$$

Here $H^{CP}(X)$ is Henry's law constant for a species $X$ in mol·Pa$^{-1}$·m$^{-3}$. $C(X)$ is the dissolved concentration in the solution with units of mol·m$^{-3}$. $P$ is the partial pressure in Pascal (1 atm ≅ 101325 Pa). Henry's law constant is also a function of temperature, which can be approximated by the van't Hoff equation.

$$\frac{dln(H^{CP})}{d(1/T)} = -\frac{\Delta H_{diss}}{R} \tag{3-2}$$

Here $T$ is the temperature in Kelvin, $R$ is the universal gas constant in J·K·mol$^{-1}$, and $\Delta H_{diss}$ is the enthalpy change in dissolution. Since there is an inverse relationship between Henry's law constant and temperature, $NH_3$ will be more soluble as temperature decreases.

## 3.2 Ocean-$NH_3$ Interaction Model

To show that $NH_3$ can accumulate in the atmosphere when life is a net source of $NH_3$ and produces enough $NH_3$ to saturate the surface sinks, we use a simplified ocean-$NH_3$ interaction model. Specifically, we study the partitioning between the atmosphere and the ocean, assuming equilibrium chemistry.

Our model has two variables: the ocean pH and the planet's total $NH_3$ reserve maintained through biological production. We assume the exoplanet has an Earth-sized ocean, roughly $1.335\times10^{21}$ L (Amante et al., 2009). Additionally, we assume there are no biological or geochemical sinks on the surface. In this case, the ocean is a reservoir. Once $NH_3$ dissolves in the ocean, it will react with water to form $NH_4^+$ and $OH^-$ ions. We calculate the dissolved ammonia concentration ($[NH_3]$) and the ammonium ion concentration ($[NH_4^+]$) by using the Henderson–Hasselbalch equation.

$$pH = pK_a + \log_{10}\left(\frac{[NH_3]}{[NH_4^+]}\right) \tag{3-9}$$

Ammonia has a $pK_a$ (acid dissociation constant at log scale) of about 9.25 (David R. Lide et al., 2005). We can derive $[NH_3]$ by solving the system of equations,

$$\frac{[NH_3]}{[NH_4^+]} = 10^{\,(pH-9.25)} \tag{3-10}$$

$$[NH_3] + [NH_4^+] = \frac{T_{NH_3}}{V_{Ocean-E}} \tag{3-11}$$

Here pH is the ocean's overall pH, $T_{NH3}$ is the planet's total $NH_3$ reserve in mol, and $V_{Ocean\text{-}E}$ is the volume of the Earth's ocean in liters. We solve [$NH_3$] and express it as a function of pH and the planet's total $NH_3$ reserve.

$$[NH_3] = \frac{T_{NH_3} \cdot 10^{\,pH}}{(1.77828 \times 10^9 + 1 \times 10^{pH}) \cdot V_{Ocean-E}} \tag{3-12}$$

Using Henry's law (Eq. 3-1), we calculate $NH_3$ partial pressure from [$NH_3$] and plot $NH_3$ volume mixing ratio as a function of pH and $T_{NH3}$.

## 3.3 Photochemistry Model

We use our photochemistry code (Hu et al., 2012) to calculate the $NH_3$ mixing ratio as a function of vertical altitude in exoplanet atmospheres. In this section, we briefly describe how we use this computational approach to explore atmospheric $NH_3$ accumulation.

Our photochemistry code (Hu et al., 2012) can simulate a wide variety of planetary atmosphere scenarios by calculating the atmosphere's steady-state chemical composition. Our full photochemistry model encodes more than 800 chemical reactions and UV photolysis of atmospheric molecules. Our model also includes both wet and dry depositions, thermal escape of O, C, H, N, and S containing molecules, surface emission, and formation and deposition of sulfur-bearing (both elemental sulfur and sulfuric acid) aerosols. The one-dimensional chemical transport model has been validated by simulating modern Earth's and Mars' atmospheric composition, matching observations of major trace gases in both scenarios. We can flexibly implement this model to simulate oxidized, oxic, and reduced atmospheres. Both ultraviolet and visible radiation in the atmosphere is computed by the delta-Eddington two-stream method. This model also includes molecular absorption, Rayleigh scattering, and aerosol Mie scattering to compute the optical depth. For applications of our photochemistry code, see (Hu et al., 2012, Seager et al., 2013(a)(b), Hu et al., 2013, Sousa-Silva et al., 2020, Zhan et al., 2021).

In this work, we consider an exoplanet orbiting an M dwarf star (M5V). We use the stellar radiation model of GJ 876 (Loyd et al., 2016; France et al., 2016). We study three different atmospheric scenarios: $N_2$-dominated weakly oxidizing atmospheres, $CO_2$-dominated highly oxidizing atmospheres, and $H_2$-dominated reducing atmospheres. For exoplanets with $N_2$-dominated, $CO_2$-dominated, and $H_2$-dominated atmospheres, the semi-major axes are 0.026 AU, 0.034 AU, and 0.042 AU, respectively. We choose the semi-major axes to ensure the surface temperature is maintained at 290 K across all scenarios. We model massive super-Earths with 10 $M_{Earth}$ and 1.75 $R_{Earth}$ because these large planets are more likely to retain an $H_2$-dominated atmosphere than Earth-mass planets due to a higher escape velocity.

In each atmosphere scenario, we consider two cases: one with $NH_3$ deposition (both dry deposition and rainout included) and one without $NH_3$ deposition (dry deposition and rainout are both set to 0). In both cases, we assume the atmospheric $NH_3$ is maintained through biological production (i.e., life is a net source of $NH_3$). The case with $NH_3$ deposition corresponds to active $NH_3$-removal sinks on the surface, where the surface is not saturated with $NH_3$. The case without $NH_3$ deposition corresponds to conditions in which life produces enough $NH_3$ to saturate the surface sinks. In this case, the ocean is a reservoir of $NH_3$, and $NH_3$ is only removed by photochemistry.

For each of the scenarios we study, we set the planet's surface pressure to 1 bar and temperature to 288 K. We divide the atmosphere into two layers: a lower convective layer and a higher radiative layer (Hu et al., 2012). We set the temperature of the convective layer to follow the dry adiabatic lapse rate. We set the stratospheric temperature in our model according to (Hu et al. 2012). We set the temperature above tropopause to 160K, 200K, and 175K for the $H_2$-dominated, $N_2$-dominated, and $CO_2$-dominated atmosphere. For an $H_2$-dominated atmosphere, we assume the stratosphere is very cold due to efficient radiative cooling from abundant $H_2$ (Birnbaum et al. 1996). Since we do not have a climate model coupled with our photochemistry code, we do not consider heating in the upper atmosphere. Hence, we assume the temperature to be constant (isothermal) in the radiative layer (Hu et al., 2012). The temperature profiles we set are consistent with significant greenhouse effects in the convective layer (Hu et al., 2012).

To study $NH_3$ as a biosignature gas, we only use 735 reactions in our simulations. Specifically, we include all the reactions mentioned in (Hu et al., 2012), except for reactions that involve more than 2 carbon atoms ($C_{>2\text{-chem}}$), high-temperature reactions (Hu et al., 2012), and $HSO_2$ thermal decay. Furthermore, we do not consider rainout and biological sinks for $H_2$, CO, $CH_4$, $N_2$, $C_2H_2$, $C_2H_4$, $C_2H_6$, and $O_2$. We make the assumptions mentioned above to facilitate robust comparison to our reference benchmark scenarios from (Hu et al., 2012). We present our simulation parameters (including the stellar spectrum input, atmospheric temperature-pressure profiles, the eddy diffusion profiles, and the planetary surface boundary information) in Supplementary Information SI.1.

## 3.4 Transmission Spectra Model and Simulated Observation

To assess future possibilities for the detection of $NH_3$ in each of the exoplanet atmospheric scenarios, we simulate transmission spectra for each scenario using our “Simulated Exoplanet Atmosphere Spectra" (SEAS) code (Zhan et al., 2021). We simulate atmospheric detection using the community JWST exposure time calculator and noise simulator Pandexo (Batalha et al., 2017) and assess the confidence in detecting $NH_3$ for each given scenario via a reduced chi-squared test.

### 3.4.1 Simulated Exoplanet Atmosphere Spectra Model (SEAS)

The SEAS transmission code takes in the output of the photochemistry code (temperature-pressure profiles and mixing ratio profiles) and calculates the optical depth along the limb path (Seager et al., 2013(a)(b); Zhan et al., 2021). The SEAS transmission code has been validated by reproducing Earth’s atmosphere spectra by comparing the SEAS simulated results with data from the Atmospheric Chemistry Experiment database (Bernath et al., 2005). Currently, our SEAS code does not have refraction. Hence our transmission spectra model works best for M dwarf planets, where refraction has a minimum impact on observations (Misra et al., 2014, Bétrémieux et al., 2014). Details of the code are described in (Zhan et al., 2021). Here we provide a very brief description of the transmission code.

We divide the entire atmosphere (from the surface to the top of the atmosphere) into layers, and the width of each layer is one atmospheric scale height. Within each layer, we assume the atmosphere is in local thermodynamic equilibrium (LTE). The stellar radiation will penetrate along each layer's limb path, which we approximate as stellar radiation passing through multiple LTE chunks. We calculate the absorption along each beam path by using the following equation.

$$A = n_{i,j} \times \sigma_{i,j} \times l_i \tag{3-13}$$

Here *A* is absorption, *n* is number density, *σ* is absorption cross-section, and *l* is the path length. The subscript *i* denotes each chunk, and subscript *j* denotes each molecule. We then compute the transmittance (*T*) of each beam by using the Beer-Lambert law. The amount of flux attenuated along each beam's path is determined by multiplying the absorbance (1-*T*) by each base chunk's scale height. Finally, we calculate the transit depth of the atmosphere by summing all the attenuated fluxes.

### 3.4.2 Simulated Exoplanet Observation

We assess an exoplanet atmospheres' detectability by using the metrics described in (Seager et al., 2013(a)(b)), (Batalha et al., 2017), and (Zhan et al., 2021). We calculate the signal-to-noise ratio (SNR) of an atmosphere using the following equation.

$$SNR = \frac{|F_{out} - F_{in}|}{\sqrt{\sigma_{F_{out}}^2 + \sigma_{F_{in}}^2}} \tag{3-14}$$

$F_{in}$ is the flux density within the absorption feature, $F_{out}$ is the flux density of the surrounding continuum of the feature, and $\sigma F_{out}$ and $\sigma F_{in}$ is the respective uncertainty. More specifically, we calculate the uncertainty using the Pandexo JWST instrumental noise calculator using the NIRSpec (G140M, G235M, G395M) and MIRI (LRS) observation modes. The two instruments combined provided a spectral coverage from 1 to about 13 µm.

To analyze the transmission detectability of weakly oxidizing ($N_2$-dominated), highly oxidizing ($CO_2$-dominated), and reducing ($H_2$-dominated) atmospheres (Section 3.3), we consider a 10 $M_{Earth}$, 1.75 $R_{Earth}$ Super-Earth orbiting a host star with a visual magnitude of 10. The central star can be either a Sun-like (G) star or a 0.2 $R_{Sun}$ M dwarf star. To observe a terrestrial exoplanet's atmosphere, we explore a total integration time between 10 transits and 100 transits (which may be impractical for G stars, but we include it for comparison).

### 3.4.3 Detection Assessment Metrics

We compare the difference in SNR to assess whether it is possible to distinguish between an $H_2$-dominated atmosphere with and without a significant $NH_3$ surface flux (Section 4.4). We assess whether the models are good fits for the simulated observational data. We use the following equation to compute the reduced chi-square statistics between the data with several different models.

$$x_v^2 = \frac{x^2}{v} = \frac{1}{v} \cdot \left[ \sum_i \frac{(O_i - C_i)^2}{\sigma_i^2} \right] \quad (3\text{-}15)$$

Here $\chi_v^2$ is the reduced chi-square, $v$ is the degree of freedom (or the number of wavelength bins), $\chi^2$ is the chi-square, $O_i$ is simulated observational data, $C_i$ is the simulated model, $\sigma_i$ is the variance (or error as calculated from Pandexo noise simulator for a specific instrument), and finally $i$ denotes each wavelength bin. We note that binning the spectra reduces the variance at the expense of reducing the degree of freedom. Due to $NH_3$ containing multiple broadband spectral features (peaks around 1.5 µm, 2 µm, 2.3 µm, 3 µm, 6 µm, and 10 µm), in practice, we find bin the spectra to a resolution of R = 10 ~ 20 is a good balance between maximizing SNR and preserving spectral information.

More specifically, we first apply a null-hypothesis test to see if the data can be explained by a straight line (i.e., no atmosphere can be detected). If the null hypothesis can be ruled out, we then compare the data for models with a variety of $NH_3$ (column-averaged mixing ratio from 0.1 ppm to 100 ppm) and assess which model best fits the data.

## 4. Results

In this section, we assess the possibility of $NH_3$ accumulation to detectable levels in various planetary scenarios. In brief, we find that $NH_3$ can theoretically accumulate the JWST detectable level in the atmosphere with a reasonable surface production flux (i.e., comparable to the terrestrial biological production of $CH_4$) only when life is a net source of $NH_3$ and produces enough $NH_3$ to saturate the surface sinks.

We first illustrate $NH_3$'s high water solubility by comparing $NH_3$'s Henry's law constant with other representative gases (Section 4.1). Using our simplified ocean-$NH_3$ interaction model and considering the atmosphere-ocean partitioning of $NH_3$, we demonstrate that $NH_3$ can theoretically accumulate if there are no $NH_3$-removal sinks on the surface (Section 4.2). Utilizing our transmission spectra model, we learn that detection of $NH_3$ is possible for exoplanets with $H_2$-dominated atmospheres using JWST if the $NH_3$ column-averaged mixing ratio reaches 5 ppm (Section 4.3). Finally, we run our photochemistry code to quantify the $NH_3$ flux needed to reach the 5 ppm detection limit under various planetary atmosphere conditions (Section 4.4).

### 4.1 $NH_3$'s High Water Solubility Limits its Ability to Accumulate in the Atmosphere

$NH_3$ is highly water-soluble, at least two orders of magnitude more soluble than other planetary atmospheric gases. $NH_3$'s high water solubility means that atmospheric accumulation of $NH_3$ is highly dependent on whether there are active $NH_3$-removal sinks on the surface (i.e., whether the surface is saturated with $NH_3$). When life does not produce enough $NH_3$ to saturate the surface sinks, rain (wet deposition) can much more efficiently remove $NH_3$ from the atmosphere than other gases, limiting $NH_3$'s ability to accumulate in the atmosphere.

We can quantitatively compare the solubility of different gases to each other based on Henry's law constants. Since Henry's law constant is a measure of solubility in water, the larger Henry's law constant, the more water-soluble the gas is. We summarize Henry's law constants for some representative gases at standard temperature and pressure (T = 298.15K and P ≅ 1atm) in Table 4.1-1 (R. Sander, 2015).

**Table 4-1.** Henry's law constant for some representative gases in water (R. Sander et al., 2015).

| Chemical species | $H^{CP}$ $[\frac{mol}{Pa \cdot m^3}]$ |
|---|---|
| Ammonia $NH_3$ | $5.90 \times 10^{-1}$ |
| Dimethyl Sulfide $CH_3SCH_3$ | $5.34 \times 10^{-3}$ |
| Methyl Chloride $CH_3Cl$ | $1.15 \times 10^{-3}$ |
| Carbon dioxide $CO_2$ | $3.36 \times 10^{-4}$ |
| Isoprene $C_5H_8$ | $1.53 \times 10^{-4}$ |
| Nitrogen dioxide $NO_2$ | $1.20 \times 10^{-4}$ |
| Phosphine $PH_3$ | $7.00 \times 10^{-5}$ |
| Methane $CH_4$ | $1.41 \times 10^{-5}$ |
| Oxygen $O_2$ | $1.30 \times 10^{-5}$ |
| Hydrogen $H_2$ | $7.75 \times 10^{-6}$ |
| Nitrogen $N_2$ | $6.43 \times 10^{-6}$ |
| Helium He | $3.83 \times 10^{-6}$ |

The high solubility of $NH_3$ and other amines (especially aliphatic amines) in water is due to their ability to form hydrogen bonds with water molecules and themselves. Even tertiary amines (three substituents are connected to N) can form hydrogen bonds between the lone pair of electrons carried by the central nitrogen atom and the H in a water molecule that carries a partial positive charge. As a result, $NH_3$ and other volatile amines are much more soluble than any previously studied biosignature gas candidates (Figure 8).

## 4.2 $NH_3$ Can Accumulate if Life Produces Enough $NH_3$ to Saturate the Surface

Using our simplified ocean-$NH_3$ interaction model, we find that $NH_3$ can theoretically accumulate to a detectable level in the atmosphere when life is a net source of $NH_3$ and produces enough $NH_3$ to saturate the surface sinks. By detectable level, we mean a volume mixing ratio of 5 ppm (see Section 4.3 for a detailed assessment of detectability of $NH_3$).

To reach this conclusion, we assess how ocean-atmosphere equilibrium $NH_3$ partial pressure changes with the planet's total $NH_3$ reserve and ocean pH. As a reminder, we assume the atmospheric $NH_3$ is maintained through biological production (as postulated by Seager et al., 2013). We also assume no $NH_3$-removal sinks on the land and in the ocean (see Section 3.2).

We find that life must produce enough $NH_3$ to maintain an $NH_3$ inventory of $6 \times 10^{18}$ mol to reach a detectable level (5 ppm) in the atmosphere. This inventory is 10,000 times that of Earth's ($1\times10^{14}$ mol). One might ask if there is enough N to support such a sizeable $NH_3$ inventory. The answer is yes, compared with Earth's total N reservoir estimated at $4\times10^{20}$ mol (Ranjan et al., 2019, Johnson et al., 2015). About ~1.5% of such a total N reservoir would be needed. Life on such planets[2] can theoretically maintain this $NH_3$ inventory by breaking down $N_2$ and $H_2$ gas to form $NH_3$, analogous to the Haber–Bosch process (Seager et al., 2013(b)).

In our model, the planet's total $NH_3$ reserve is limited by the planet's total N reservoir. Unfortunately, the upper bound of the total N reservoir cannot be known for an exoplanet. Here, we assume the planet has an Earth-sized total N reservoir of $4\times10^{20}$ mol. In an extreme situation where life manages to convert almost all of the planetary N reservoir into $NH_3$, the $NH_3$ volume mixing ratio can reach 0.5% (Figure 3b). In this case, $NH_3$ becomes one of the major chemical species in the atmosphere, and the ocean becomes basic. However, if life cannot produce a substantial amount of $NH_3$ and only maintain an Earth-like $NH_3$ inventory (~$10^{14}$ mol), $NH_3$ will not be detectable regardless of the ocean's pH levels (Figure 3a).

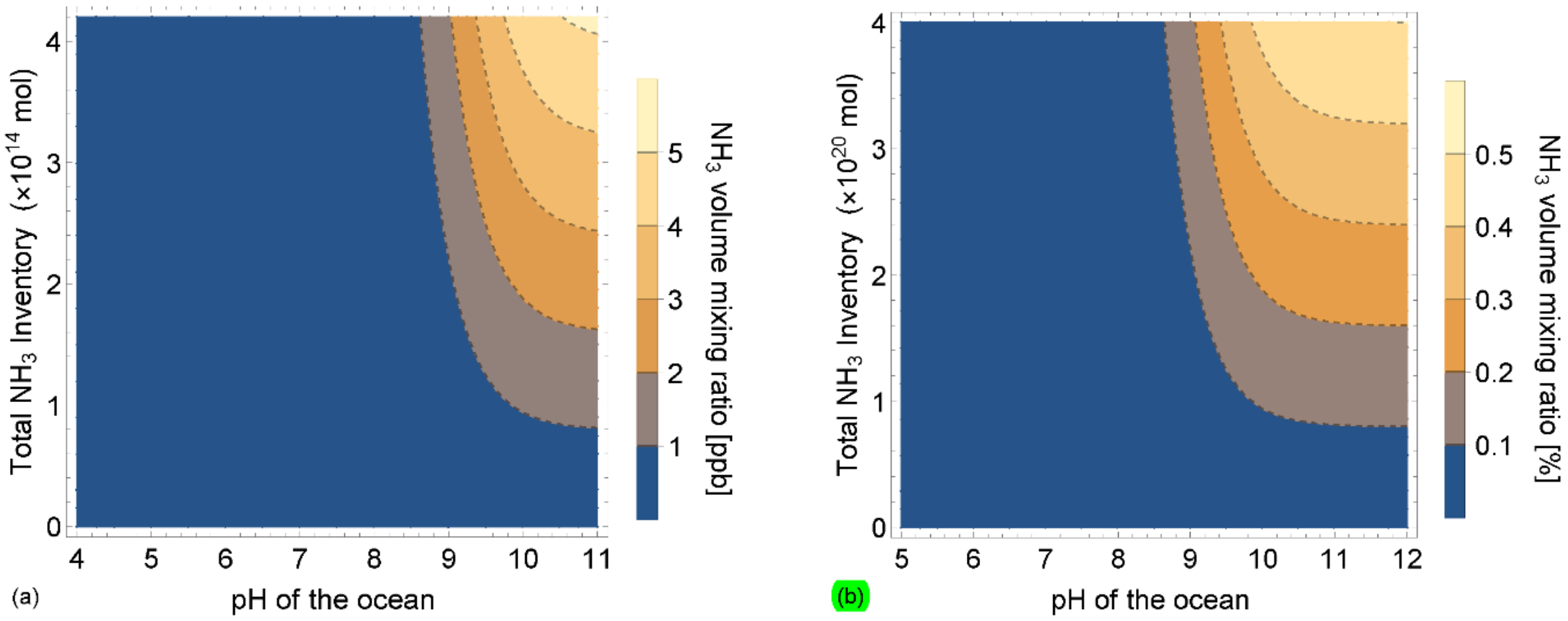

**Figure 3**. Equilibrium volume mixing ratio of $NH_3$ as a function of the planet's total $NH_3$ reserve and ocean pH. The x-axis is the ocean pH, and the y-axis is the planet's total $NH_3$ reserve in mol. The contours are $NH_3$ volume mixing ratios. We assume the planet has an Earth-sized total N reservoir ($4\times10^{20}$ mol) and an Earth-sized ocean ($1.335\times10^{21}$ L). (a): If life cannot produce a substantial amount of $NH_3$ and only maintain an Earth-like $NH_3$ inventory (~$10^{14}$ mol), the $NH_3$ volume mixing ratio is extremely low regardless of the ocean pH level ($4 \leq pH \leq 11$). (b): If life manages to convert almost all of the planetary N reservoir into $NH_3$ (~$10^{20}$ mol), $NH_3$ will become one of the major chemical species in the atmosphere.

## 4.3 Detectability of $NH_3$ in Transmission Spectra

We study the detectability of $NH_3$ in transmission spectra using our 'Simulated Exoplanet Atmosphere Spectra' (SEAS) code. For exoplanets with $H_2$-dominated atmospheres orbiting M dwarf stars (M5V), we find that a minimum of about 5 ppm column-averaged mixing ratio is needed to detect $NH_3$ with JWST, considering $NH_3$'s 1.5 μm feature and a 10 ppm JWST systematic noise floor (Figure 4). Furthermore, we find that exoplanets with high molecular-weight atmospheres (i.e., $CO_2$-dominated and $N_2$-dominated atmospheres) and those orbiting Sun-like stars have atmosphere signals too weak to be detected by JWST.

---

[2] We note that $H_2$-dominated atmospheres are not detrimental for life and that life can survive and thrive in an $H_2$-dominated environment (Seager et al., 2020).

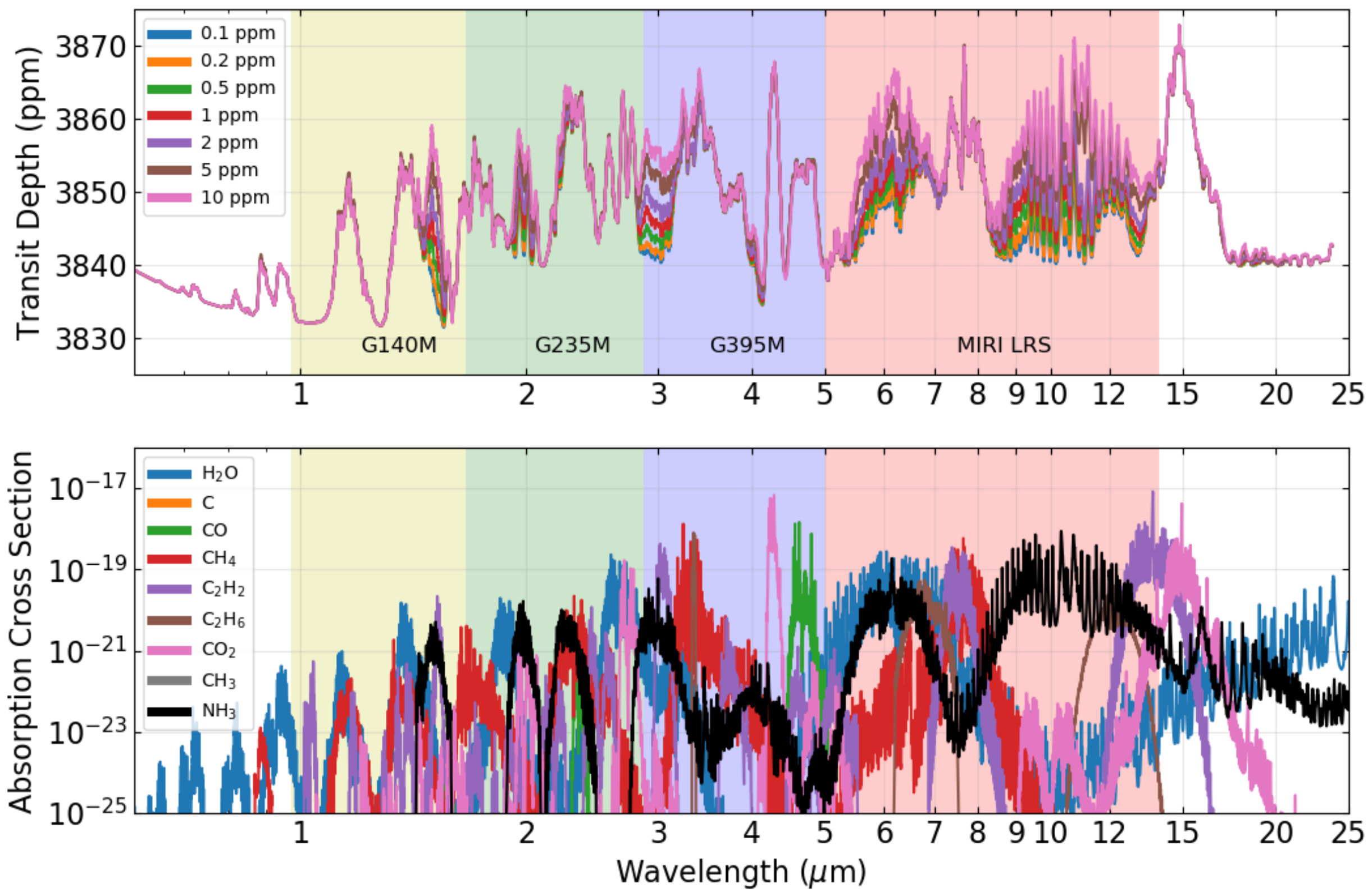

**Figure 4**. Simulated spectra of an exoplanet with an $H_2$-dominated atmosphere transiting an M5V star for a range of $NH_3$ column-averaged mixing ratios. The y-axis shows transit depth (ppm), and the x-axis shows wavelength (μm). The spectra are simulated from 0.3 - 23 μm, covering the wavelength span of most of JWST's observation modes. The yellow, green and blue region shows the spectral coverage of NIRSpec, and the red region shows that of MIRI LRS. We simulate the spectra with varying $NH_3$ column-averaged mixing ratios from 0 (no $NH_3$) to 10 ppm. We calculate the $NH_3$ mixing ratio as a function of vertical altitude using our photochemistry code (Section 3.3). At lower surface mixing ratios, the $NH_3$ spectral features are not prominent as $NH_3$ is mostly concentrated near the surface and rapidly decays as a function of altitude. Increasing the abundance of $NH_3$ amplifies the spectral feature of $NH_3$, as more $NH_3$ can accumulate in the upper part of the atmosphere.

For exoplanets with high molecular-weight atmospheres (i.e., $CO_2$-dominated and $N_2$-dominated atmospheres), the transit depth resulting from the small atmosphere scale height is too small compared to the assumed 10 ppm JWST observational noise floor. Similarly, for exoplanets orbiting Sun-like stars, the transit depth (around 2 ~ 3 ppm for $H_2$-dominated atmospheres) is too small given the size of the host star. To support this argument, we conduct a null hypothesis test. We check whether our model is a better fit than a flat line. We find that only exoplanets with $H_2$-dominated atmospheres orbiting M5V stars pass this test[3].

To demonstrate that a minimum of about 5 ppm is needed for $NH_3$ to be detected, we simulate JWST observations using the Pandexo JWST noise simulator with a detection noise floor of 10 ppm and a realistic noise model for JWST instrumentation (Batalha et al., 2017). We compare a simulated spectrum with 5 ppm of $NH_3$ and a spectrum with no $NH_3$ (Figure 5). We find that the spectrum with 5 ppm of $NH_3$ has spectral features 10 ~ 40 ppm larger than the spectrum with no $NH_3$. The inclusion of detection noise is the main reason why our 5 ppm detection limit is higher than the 0.1 ppm value stated in (Seager et al., 2013).

[3] Except for TRAPPIST-1 planets (details see Supplementary Information SI.9)

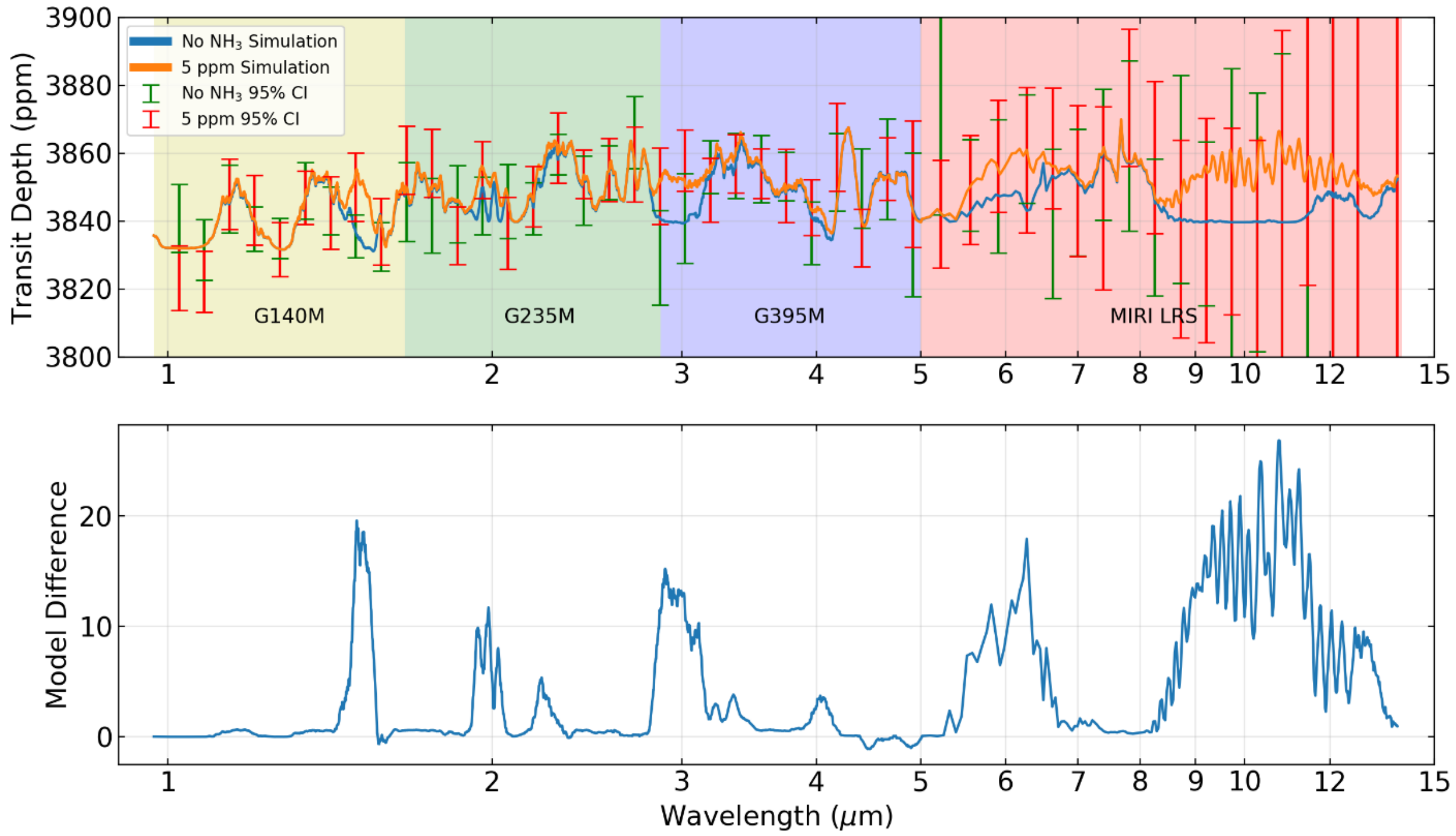

**Figure 5. Upper Panel**: Simulated JWST exoplanet atmosphere observation for a 10 $M_{Earth}$, 1.75 $R_{Earth}$ super-Earth with an $H_2$-dominated atmosphere transiting an M5V star given 20 transit per instrument (80 transits in total), and comparing a model with no surface $NH_3$ (blue line) and a model with a column average mixing ratio of 5 ppm (orange line), which we also show in Figure 4. The y-axis shows transit depth (ppm), and the x-axis shows wavelength (µm). The simulated observation spans the wavelength range of the NIRSpec and MIRI instruments. The error bars are 95% confidence interval for each model with 5 ppm $NH_3$ (red) and green for no $NH_3$ (green) uniformly binned to a spectral resolution of R=10. We show that the difference between the two models achieves statistical significance within each instrument, indicating a confident simulated detection of $NH_3$. **Lower Panel**: Model difference in ppm between the two simulated spectra showing the spectral feature of $NH_3$ peaks around 1.5 µm, 2 µm, 3 µm, 6 µm, and 10 µm. Negative values denote that the increase of $NH_3$ caused a decrease of $CH_4$ in the atmosphere.

Even though we find that a minimum of about 5 ppm is needed for $NH_3$ to be detected, with one JWST instrument (e.g., NIRSpec) and one mode (i.e., G395M, which encompasses the $NH_3$'s 3 µm feature), constraining the amount of $NH_3$ will be challenging. A better quantification is possible with more modes covered. We have analyzed an ideal case with observations of all $NH_3$ spectral features in the wavelength range of 1.6 to 10 µm (three modes of NIRSPEC and MIRI LRS, i.e., 80 transits). In this case, we can further constraint the $NH_3$ mixing ratio to 1 ~ 10 ppm.

For illustration purposes, we also show an instance of a simulated observation of an exoplanet with $H_2$-dominated atmospheres transiting an M5V star, with both a reasonable number of transits (20 transits; Figure 6a) and the maximum theoretical observation time possible (200 transits; Figure 6b). The maximum observation time means considering nearly every visible transit for an orbital period corresponding to the habitable zone of an M5V star (i.e., one transit every 10 days for five years based on Kepler's 3rd law). While such a 200 transit-observation is unrealistic (due to the competitive nature of JWST), Figure 6 illustrates what quality of data 200 transits would provide.

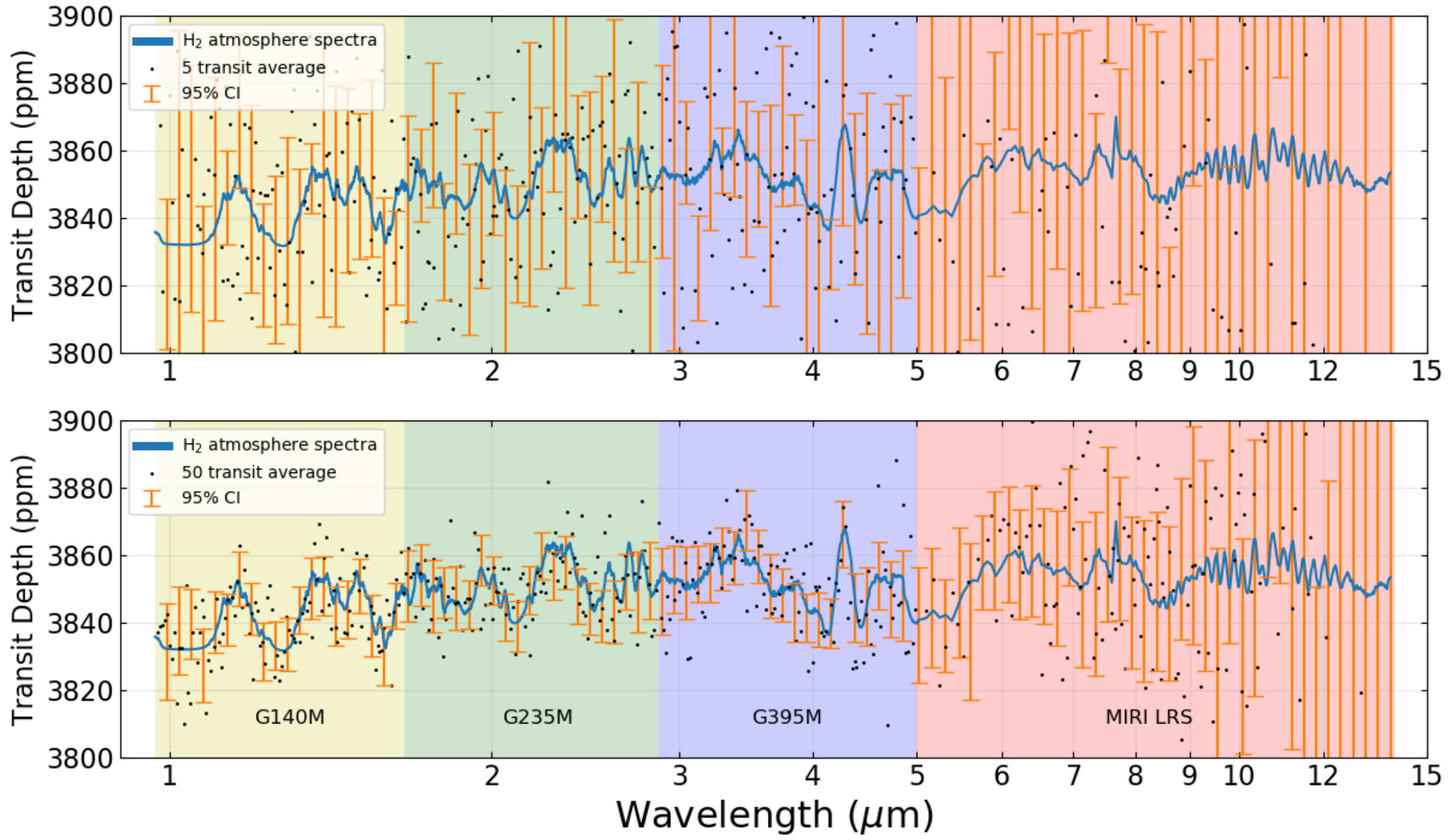

**Figure 6.** Comparison between 20 and 200 transits for simulated JWST exoplanet atmosphere observations for a 10 $M_{Earth}$, 1.75 $R_{Earth}$ super-Earth with $H_2$-dominated atmosphere transiting an M5V star. Similar to Figure 5, the model shows a planet with a 5 ppm column-averaged mixing ratio of $NH_3$. **Upper panel**: simulation for 20 transit observations (5 transits per instrument modes) and R = 20. The y-axis shows transit depth (ppm), and the x-axis shows wavelength (µm). The simulated observation focuses on the wavelength span of NIRSpec G140M (yellow), G235M (green), G395M (blue), and MIRI LRS (red), respectively. The blue line is the simulated transmission spectra, and the black dots are the average of all transits. The orange error bars represent the 95% confidence interval of the observation. **Lower panel**: same as the upper panel but for 200 transits (50 per mode). This simulation assumes the maximum observation time possible with JWST (200 transits, spanning across 5 years), and spectral features of various molecules can be characterized to high confidence.

## 4.4 Variability of $NH_3$ Atmospheric Accumulation in Various Planetary Scenarios

We use our photochemistry model to simulate the $NH_3$ mixing ratio as a function of vertical altitude in various planetary atmosphere scenarios. We first demonstrate that $NH_3$ can accumulate in the atmosphere with a reasonable surface production flux only when life produces enough $NH_3$ to saturate the surface sinks (Section 4.4.1). We also show the effect of the $NH_3$ surface deposition on atmospheric $NH_3$ mixing ratio given a fixed $NH_3$ surface production flux (Section 4.4.2).

### 4.4.1 Conditions Required for $NH_3$ to Accumulate with a Reasonable Surface Production Flux

We find that $NH_3$ can accumulate to the 5 ppm JWST-detectable level in the atmosphere with a reasonable surface production flux only if life is a net source of $NH_3$ and produces enough $NH_3$ to saturate the surface sinks. In this case, there are no $NH_3$-removal sinks on the surface, and $NH_3$ is only removed by photochemistry (i.e., no dry deposition or rainout).

Specifically, for an exoplanet with $H_2$-dominated atmospheres orbiting an M dwarf star, when the surface is saturated with $NH_3$ (i.e., there are no $NH_3$-removal sinks on the surface), the required biological surface flux to reach 5 ppm is on the order of $10^{10}$ molecules cm$^{-2}$ s$^{-1}$, comparable to the terrestrial biological production of $CH_4$. However, when the surface is unsaturated with $NH_3$, due to additional sinks present on the surface, life would have to produce $NH_3$ at surface flux levels on the order of $10^{15}$ molecules cm$^{-2}$ s$^{-1}$ (~$4.5\times10^6$ Tg year$^{-1}$). This value is roughly 20,000 times greater than the biological production of $NH_3$ on Earth and about 10,000 times greater than Earth's $CH_4$ biological production.

**Table 4-1.** Simulated mixing ratios and surface fluxes for exoplanets with $H_2$-dominated, $CO_2$-dominated, and $N_2$-dominated atmospheres orbiting M dwarf stars (M5V).

| Atmospheric scenarios | $NH_3$ column-averaged mixing ratio | $NH_3$ surface flux needed [molecules cm$^{-2}$ s$^{-1}$] | |
|---|---|---|---|
| | | With $NH_3$ deposition | Without $NH_3$ deposition |
| $H_2$-dominated | $5.0 \times 10^{-6}$ (5 ppm) | $6.40 \times 10^{15}$ | $1.44 \times 10^{10}$ |
| $CO_2$-dominated | $5.0 \times 10^{-6}$ (5 ppm) | $3.60 \times 10^{14}$ | $8.49 \times 10^{8}$ |
| $N_2$-dominated | $5.0 \times 10^{-6}$ (5 ppm) | $7.10 \times 10^{14}$ | $6.77 \times 10^{10}$ |

As a reminder, we assume the atmospheric $NH_3$ is maintained through biological production in our simulations. The case with $NH_3$ deposition corresponds to active $NH_3$-removal sinks on the surface, where the surface is not saturated with $NH_3$. The case without $NH_3$ deposition corresponds to conditions in which life produces enough $NH_3$ to saturate the surface sinks (see Section 3.3). We include the $N_2$- and $CO_2$-dominated scenario for completeness even though we show exoplanets with high molecular-weight atmospheres (i.e., $CO_2$-dominated and $N_2$-dominated atmospheres) have atmosphere signals too weak to be detected by JWST (see Section 4.3 for more details).

Additionally, we find that the dominant photochemical removal pathway for $NH_3$ is direct photolysis, followed by reactions with OH radicals. Except in $CO_2$-dominated atmospheres, the second dominant removal pathway is reactions with excited O radicals. We compile the top three loss rates for $NH_3$ in Table S5-1 (see Supplementary Information SI.5).

To test our photochemistry results' robustness, we perform various sensitivity tests (including the presence of a cold trap, the choice of eddy diffusion coefficient, and Henry's law constant for $NH_3$). We find that our results are not sensitive to any of these variables (see Supplementary Information SI.6). Furthermore, we find that $NH_3$ is insufficiently abundant to condense in our simulated atmospheres (see Supplementary Information SI.7).

### 4.4.2 Effects of $NH_3$ Surface Deposition on Atmospheric $NH_3$ Mixing Ratio

Given a fixed $NH_3$ surface production flux, the presence of $NH_3$ surface deposition has a significant effect on the atmospheric $NH_3$ mixing ratio. Specifically, the atmospheric mixing ratio of $NH_3$ is several orders of magnitude lower with deposition than without (Figure 7).

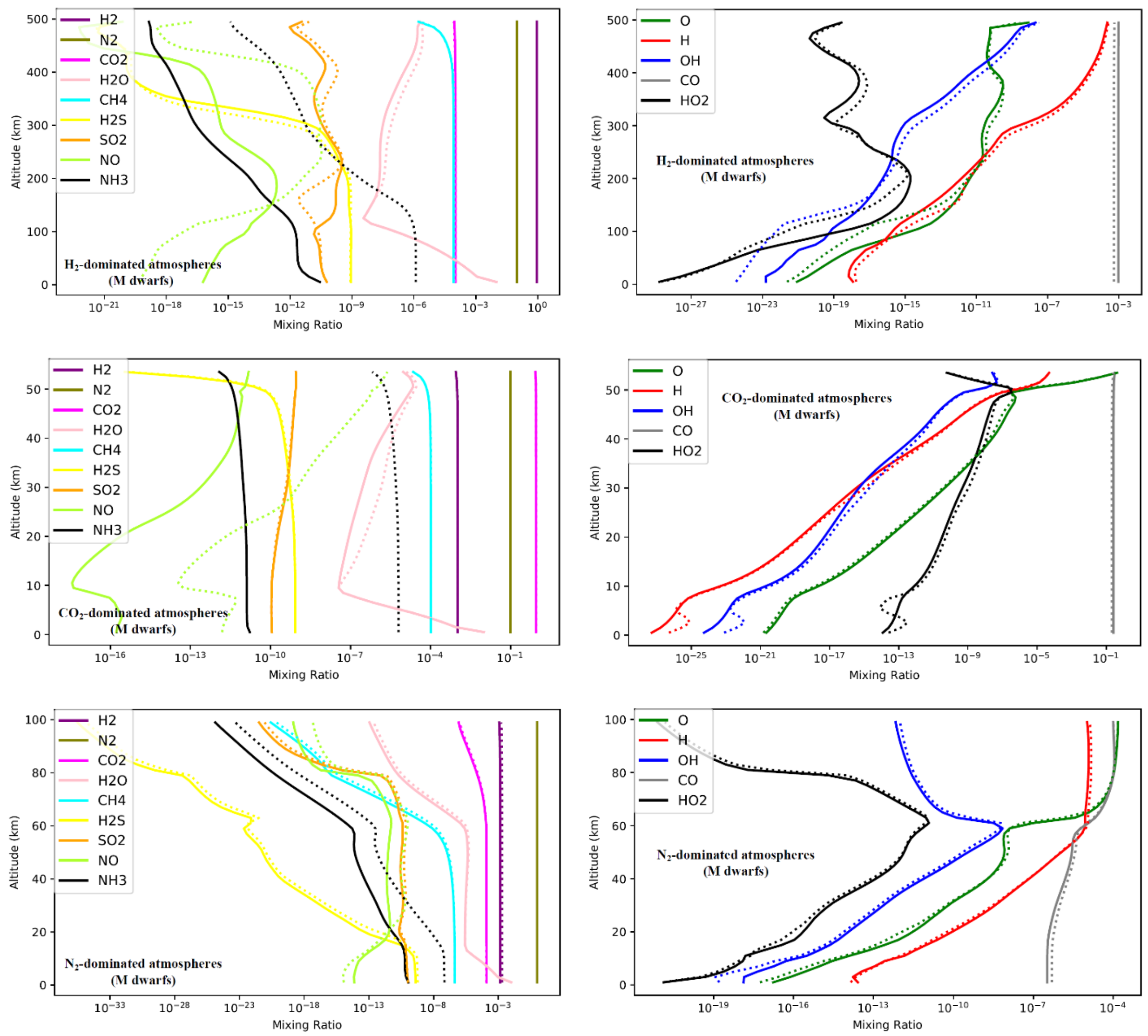

**Figure 7**. The volume mixing ratio of various atmospheric species on planets with $H_2$-dominated, $CO_2$-dominated, and $N_2$-dominated atmospheres orbiting active M dwarfs. The $NH_3$ surface production flux is $1.0\times10^{10}$ molecule·cm$^{-2}$·s$^{-1}$ for the $H_2$- and $N_2$-dominated atmospheres. For the $CO_2$-dominated case, the flux is $1.0\times10^{9}$ molecule·cm$^{-2}$·s$^{-1}$. The y-axis shows altitude in km, and the x-axis shows the mixing ratio. Note that each row of figures has a different x-axis scale. Each color denotes one particular species. The dotted lines for $NH_3$ deposition absent, and solid lines for $NH_3$ deposition present. The figures in the left panel show molecular concentrations, and the figures in the right panel show radical concentrations under different atmospheric scenarios. In each of the three atmospheric scenarios we simulate, the solid black curve ($NH_3$ mixing ratio with deposition) is shifted further to the left (smaller mixing ratio) compared to the dotted black curve ($NH_3$ without deposition) since $NH_3$ deposition suppresses atmospheric $NH_3$ concentration. Note that the solid curves for species other than $NH_3$ show the effects of $NH_3$ deposition on the atmospheric concentration of other species.

When life cannot produce a substantial amount of $NH_3$, the surface remains unsaturated with $NH_3$. Hence, surface deposition (both rainout and dry deposition) effectively removes $NH_3$ from the atmosphere. If life produces enough $NH_3$ to saturate the surface, surface deposition no longer plays a role, and $NH_3$ is only removed by photochemistry.

The effect of surface deposition on $NH_3$ mixing ratio differs between the $H_2$-, $N_2$- and $CO_2$-dominated atmospheres. Specifically, we find that rainout (i.e., wet deposition) is most effective in $H_2$-dominated atmospheres compared to $N_2$- or $CO_2$-dominated atmospheres. The difference in atmospheric $NH_3$ mixing ratio with and without surface deposition[4] can reach six orders of magnitude. The effect of rainout is most significant in $H_2$-dominated atmospheres due to the thermodynamics of $H_2$. The tropospheric lapse rate is lower in $H_2$-dominated atmospheres than either $CO_2$- or $N_2$-dominated atmospheres, driving higher temperatures, higher water content, and higher rainout rates.

Even though surface deposition (particularly rainout) is not as effective in $CO_2$-dominated atmospheres as $H_2$-dominated atmospheres, surface deposition can make a massive difference in the $NH_3$ mixing ratio in $CO_2$-dominated atmospheres. In the absence of surface deposition, $NH_3$ concentrations are the lowest in the relatively oxidizing $CO_2$-dominated atmospheres (compared to the $N_2$- and $H_2$-dominated atmospheres), as one might naively expect on the grounds of simple thermodynamics. With the presence of surface deposition, $NH_3$ concentrations are the highest in the $CO_2$-dominated atmospheres. The reason is that the wet deposition rate decreases as temperature and atmospheric water content decreases (Giorgi & Chameides 1985; Hu et al., 2012, Equation 21). The relatively high lapse rate of $CO_2$-dominated atmospheres suppresses both temperature and atmospheric water content, reducing rainout and enhancing $NH_3$ concentrations relative to the $N_2$- and $H_2$-dominated atmospheric scenarios.

## 5. Discussion

We first compare $NH_3$'s solubility in water with other atmospheric gases, particularly several previously studied biosignature gases (Section 5.1). We next discuss how horizontal atmospheric transport might limit $NH_3$ accumulation above land (Section 5.2). We also discuss other minor sources and sinks of ammonia, including $NH_3$ production by lightning (Section 5.3) and additional ocean-related sinks for dissolved $NH_3$ and $NH_4^+$ ions (Section 5.4). We discuss amines viability as biosignature gases by proxy with $NH_3$ (Section 5.5 and Section 5.6). Additionally, we briefly discuss $NH_3$ induced hazes (Section 5.7), $NH_3$'s greenhouse effect (Section 5.8), and $NH_3$ detectability with future telescopes (Section 5.9). We end our discussion with a brief comparison of the significance of atmospheric $NH_3$ in mini-Neptunes vs. super-Earths (Section 5.10).

### 5.1 Ammonia's High Water Solubility Compared to Other Atmospheric Gases

We compare ammonia's solubility to that of 27 other gases, a list that includes common atmosphere gases, biosignature gases, and a group of volatile amines (Figure 8).

[4] $NH_3$ dry deposition velocity is the same for all the scenarios.

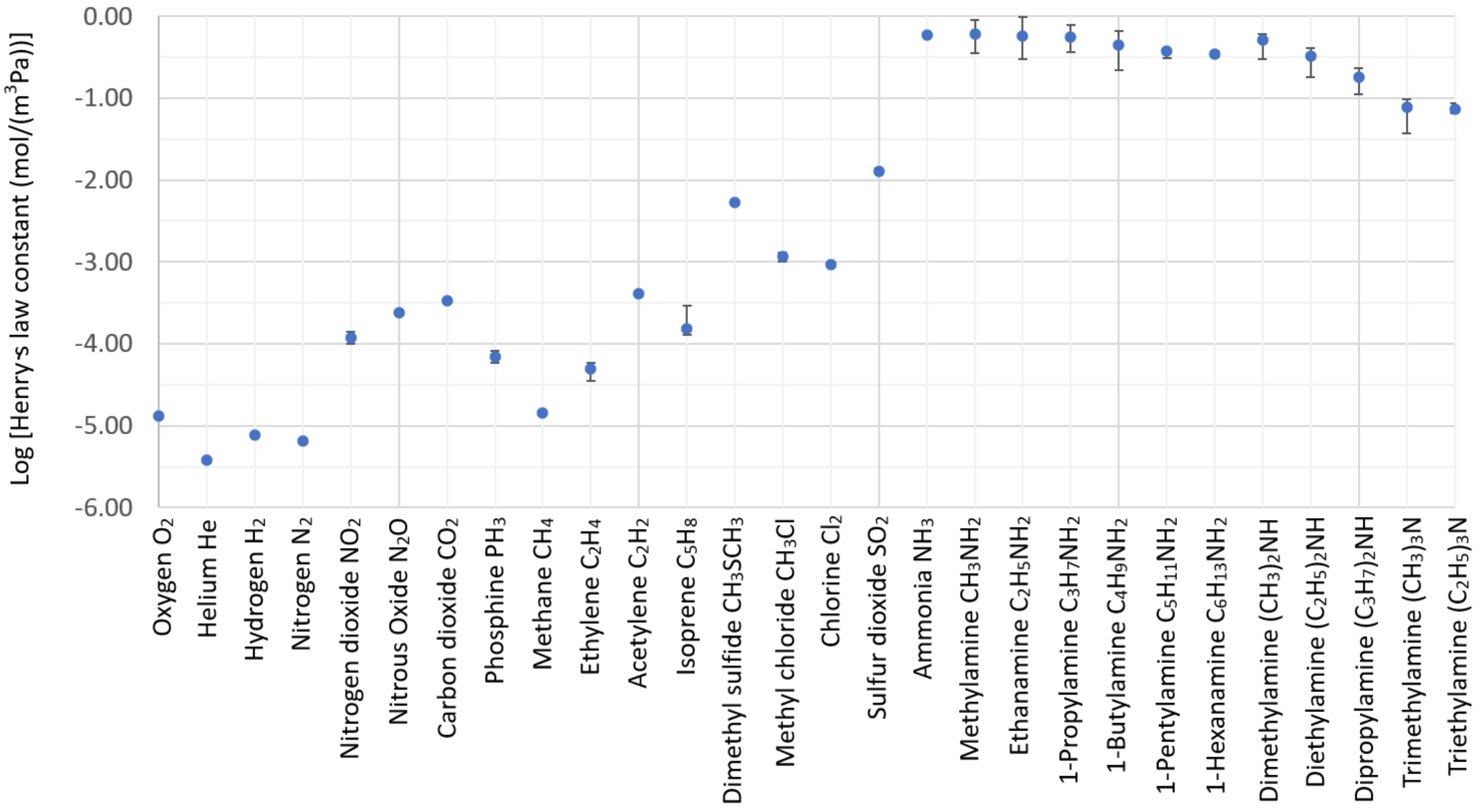

**Figure 8**. The solubility of common molecules on a log scale. The x-axis shows the chemical species' name, and the y-axis shows Henry's law constant on a log scale. Ammonia and other amines are at least two orders of magnitude more soluble than other chemicals in the list, including several biosignature gas candidates that have already been studied.

In contrast to ammonia and volatile amines, previously studied biosignature gases such as isoprene (Zhan et al., 2021), DMS (Domagal-Goldman et al., 2011, Seager et al., 2012, and Arney et al., 2018), $CH_4$ (Dlugokencky et al., 2011), $N_2O$ (Tian et al., 2015), $CH_3Cl$ (Segura et al., 2005) and phosphine (Sousa-Silva et al., 2019) have very low solubilities in water. Therefore, molecule water solubility has not needed to be emphasized as a reservoir in exoplanet atmosphere characterization studies.

Sulfur dioxide ($SO_2$) has a much higher solubility in water than other common atmospheric gases (but not as high as $NH_3$). Many studies show that both $SO_2$ molecules and their associated aerosols ($H_2SO_4$-$H_2O$) can potentially be detected in exoplanets' atmospheres (Hu et al., 2013, Lincowski et al., 2018, Loftus et al., 2019). However, Earth does not sustain a detectable level of sulfur aerosols in the atmosphere, possibly due to the presence of a global ocean (McCormick et al., 1995 and Loftus et al., 2019). Furthermore, it has been shown that $SO_2$ accumulation in an Earth-like environment requires an unreasonably high source flux (Hu et al., 2013). Therefore, (Loftus et al., 2019) propose that detection of atmospheric $SO_2$ (or sulfate haze) can infer the lack of surface water oceans on a rocky exoplanet.

## 5.2 Horizontal Atmospheric Transport Limits $NH_3$ Accumulation above Land

Horizontal wind transport might limit $NH_3$ accumulation above land when the planet's surface is unsaturated with $NH_3$ (i.e., there are $NH_3$ biological or geochemical sinks on the surface). In this case, if land-based life produces $NH_3$, it can be transported over the ocean, where $NH_3$ can rain out and dissolve in the ocean, thereby removing $NH_3$ from the atmosphere. Since Earth's $NH_3$ is produced by land-based life (soil bacteria (Section 2.1), horizontal wind transport is important.

So far, our models have neglected wind transport because they are limited to one dimension (the vertical dimension). Horizontal wind transport is a sink through which $NH_3$ is removed from the atmosphere above land. An improved model would consider an $NH_3$ land-ocean transport cycle (Figure 9) that includes horizontal winds. After $NH_3$ is produced from land surfaces, it will either be transported horizontally by wind or move up to the upper atmosphere, where $NH_3$ will be photochemically destroyed. Once entering the atmosphere above the ocean, ammonia can rain out, followed by quick diffusion into the ocean due to its high solubility. As a result, wind transport limits $NH_3$'s ability to accumulate to a detectable level in the atmosphere. A numerical model for such atmospheric wind transport of $NH_3$ is very challenging since it's extremely sensitive to the geophysical conditions we assume for the exoplanets.

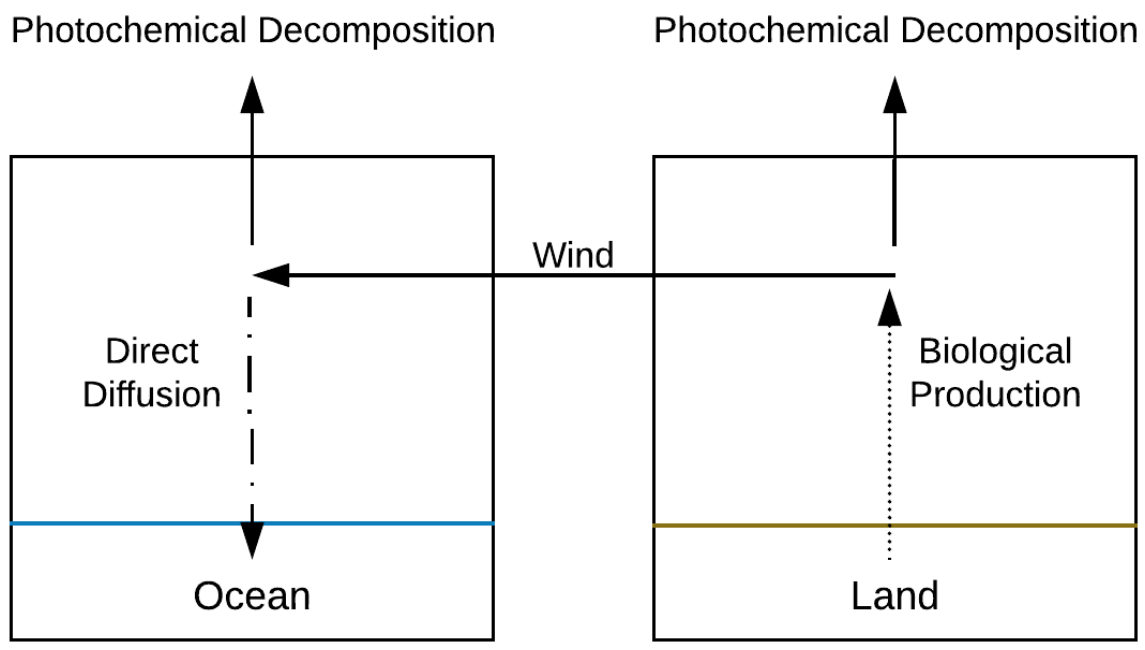

**Figure 9**. A simplified $NH_3$ land-ocean transport cycle. We assume that the planet's surface (land and ocean) is unsaturated with $NH_3$, meaning there are $NH_3$ biological or geochemical sinks on the surface. Right panel: Ammonia is biologically produced from land surfaces and enters the atmosphere. Ammonia leaves the land area by traveling to the upper atmosphere, where it is photochemically destroyed. $NH_3$ can also be carried away by horizontal wind transport. Left panel: Once entering the atmosphere above the ocean, ammonia is removed through direct diffusion into the ocean, where $NH_3$ readily dissolves due to its high solubility. $NH_3$ can also fall back to the ground with rain, which is not shown in this figure.

### 5.3 Implausibility of $NH_3$ Formation by Lightning

In principle, the formation of $NH_3$ from $N_2$ is possible during lightning strikes. However, the reduction of $N_2$ to $NH_3$ by lightning is unlikely in oxidized atmospheres due to the very low concentration of H-containing reductants (e.g., $H_2$). N species' well-studied formation by lightning strikes on Earth ($N_2$-$O_2$ atmosphere) favors the formation of nitrates and nitrites, not the thermally less stable reduced forms of N like ammonia (Ardaseva et al., 2017; Mancinelli and McKay 1988; Rakov and Uman 2003). Similarly, in oxidized $N_2$-$CO_2$ atmospheres, lightning-induced $N_2$ transformation primarily leads to HNO but not $NH_3$ (Navarro-González et al., 2001; Holloway et al., 2002; Hawtof et al., 2019).

In an $H_2$-dominated atmosphere, $NH_3$ may form through lightning by breaking apart the $N_2$ bond and subsequent recombination of N atoms with H. Unfortunately, there are no laboratory studies on the formation of ammonia in $N_2$-$H_2$ atmospheres. Several studies focusing on the lightning-induced formation of $PH_3$ conclude it is possible, though extremely low efficiency. Specifically, simulated lightning discharges in laboratory conditions produce traces of $PH_3$, at very low efficiency, from discharges onto phosphate salt solutions (Glindemann et al., 1999; Glindemann et al., 2004). By analogy with $PH_3$, we infer that lightning is not likely an efficient source of $NH_3$ in exoplanet atmospheres.

### 5.4 Plausible Ocean-related Sinks of $NH_3$ and $NH_4^+$

There are active $NH_3$-removal sinks on the surface (including land and ocean) that can maintain a surface unsaturated with $NH_3$. We summarize three chemical processes that can remove dissolved $NH_3$ or $NH_4^+$ ions from the ocean: abiotic and biotic ammanox, marine photochemistry, and the formation of $NH_4$-containing sediments and minerals in the ocean.

The first process, called anammox (anaerobic ammonium oxidation), is a process that combines $NH_4^+$ and $NO_2^-$ to form $N_2$ and water. Such a process can occur biotically with marine anammox bacteria such as species belonging to the genus *Scalindua* (Jetten et al., 2009). Abiotic anammox is also possible. The rate of abiotic reaction of $NH_3$ with nitrite is sensitive to pH and temperature, with the reaction proceeding substantially faster at low pH and high temperature (Nguyen et al., 2003). $NH_3$ may also react with $NO_3^-$, but this reaction is slower than reaction with $NO_2^-$, by analogy with similar reduction reactions (Ranjan et al., 2019).

The second process is marine photochemistry. UV radiation in the wavelength range of 300-400 nm can effectively penetrate the first few centimeters of the surface ocean water[5] before it is attenuated (Fleischmann 1989). Therefore, dissolved $NH_3$ and $NH_4^+$ ions might be photochemically destroyed by the incident UV light close to the ocean surface. The bond dissociation energy for an N-H single bond is about 314 $kJ \cdot mol^{-1}$ (T. L. Cottrell, 1966). Photons with wavelengths between 300 and 400 nm have 299 ~ 399 $kJ \cdot mol^{-1}$ of energy. Therefore, it is not implausible for UV to break the N-H bond, effectively removing $NH_3$ or $NH_4^+$ ions from the ocean surface.

The third process is the formation of $NH_4$-containing sediments and minerals in the ocean, whereby dissolved $NH_4^+$ ions are deposited into the lithosphere. There are a wide variety of $NH_4$-containing minerals in nature (Table 5-1). $NH_4$-containing minerals like Lecontite, Ammonian fluorapophyllite, and Tobelite can form by substitution of ammonium for potassium. High concentrations of $NH_4^+$ ions have been detected in the mineralizing fluids at the bottom of the ocean, specifically around the southwest pacific regions (Ridgway et al., 1990). Organic matter embedded in the sediments and rocks can break down to yield $NH_4^+$. Since $NH_4$ can easily replace K, Na, and other alkali metals in crystal lattices, K-containing rocks and sediments are particularly susceptible to the formation of $NH_4$ haloes and incursions (Ridgway et al., 1990). Both surface and subsurface rocks can have high $NH_4$ content (Ridgway et al., 1990). As a result, dissolved $NH_3$ and $NH_4^+$ ions can, in principle, be removed from the ocean environment through this mineralogical process.

[5] The UV irradiance at the surface of an anoxic planet can be substantially greater than that of present-day Earth, due to the much more efficient penetration of the shorter UV wavelengths (Cockell, 1999).

**Table 5-1.** Naturally occurring $NH_4$-containing minerals (Reference: Holloway et al., 2002).

| Name of the mineral | Chemical formula |
|---|---|
| Ammoniojarosite | $(NH_4)Fe_3^+(SO_4)_2(OH)_6$ |
| Boussingaultite | $(NH_4)_2Mg(SO_4)_2 \cdot 6H_2O$ |
| Letovicite | $(NH_4)_3H(SO_4)_2$ |
| Mascagnite | $(NH_4)_2SO_4$ |
| Sal-ammoniac | $(NH_4)Cl$ |
| Mundrabillaite | $(NH_4)_2Ca(HPO_4)_2 \cdot H_2O$ |
| Ammonioleucite | $(NH_4)AlSi_2O_6$ |
| Lecontite | $(NH_4,K)Na(SO_4) \cdot 2H_2O$ |
| Ammonian fluorapophyllite | $(NH_4,K)Ca_4Si_8O_{20}(F,OH) \cdot 8H_2O$ |
| Tobelite | $(NH_4,K)Al_2(Si_3Al)O_{10}(OH)_2$ |

## 5.5 Volatile Amines are Good Biosignature Gases by Proxy with $NH_3$

Volatile amines have similar solubilities and reactivities to $NH_3$ and hence are also suitable biosignature gases. Only when life produces enough volatile amines that can saturate the surface sinks can volatile amines accumulate in the atmosphere. Otherwise, volatile amines, just like $NH_3$, can be removed from the atmosphere due to their high water solubility and high bio-useability.

Amines can be considered $NH_3$ derivatives, where an organic functional group replaces at least one H atom. There is some difference in solubility amongst volatile amines. In general, amine solubility decreases with an increasing length or number of hydrocarbon chains. Furthermore, like $NH_3$, volatile amines will form basic solutions in water (see SI.2).

The reactivity of volatile amines is also similar to $NH_3$ (see SI.3), further allowing us to use $NH_3$ as a proxy for volatile amines. Here, we compare radical reaction rates[6] for volatile amines to $NH_3$ in NIST (Manion et al., 2008). The few volatile amines with radical reaction rates at room temperature have reaction rates two to three orders of magnitude higher than that of $NH_3$ (see SI.3). As a result, if $NH_3$ cannot accumulate in the atmosphere unless life produces a substantial amount of $NH_3$ that can saturate the surface sinks, neither can other volatile amines under similar planetary conditions.

## 5.6 IR Absorbance Spectra of Amines

We can organize amines into three groups: primary, secondary, and tertiary amines, despite the amines' wide variety of molecular structures. Amines in each group will show similar features in their IR spectra. We use IR absorbance data from NIST to plot the spectra. We find 15 primary amines, 9 secondary amines, and 2 tertiary amines with available absorbance data in NIST. We overlay them onto a single graph to demonstrate the presence of shared IR features (Figure 10).

[6] H radicals do not destroy $NH_3$ or other volatile amines.

We present the details of those amines and their raw (unnormalized) plots in Supplementary Information (SI.4).

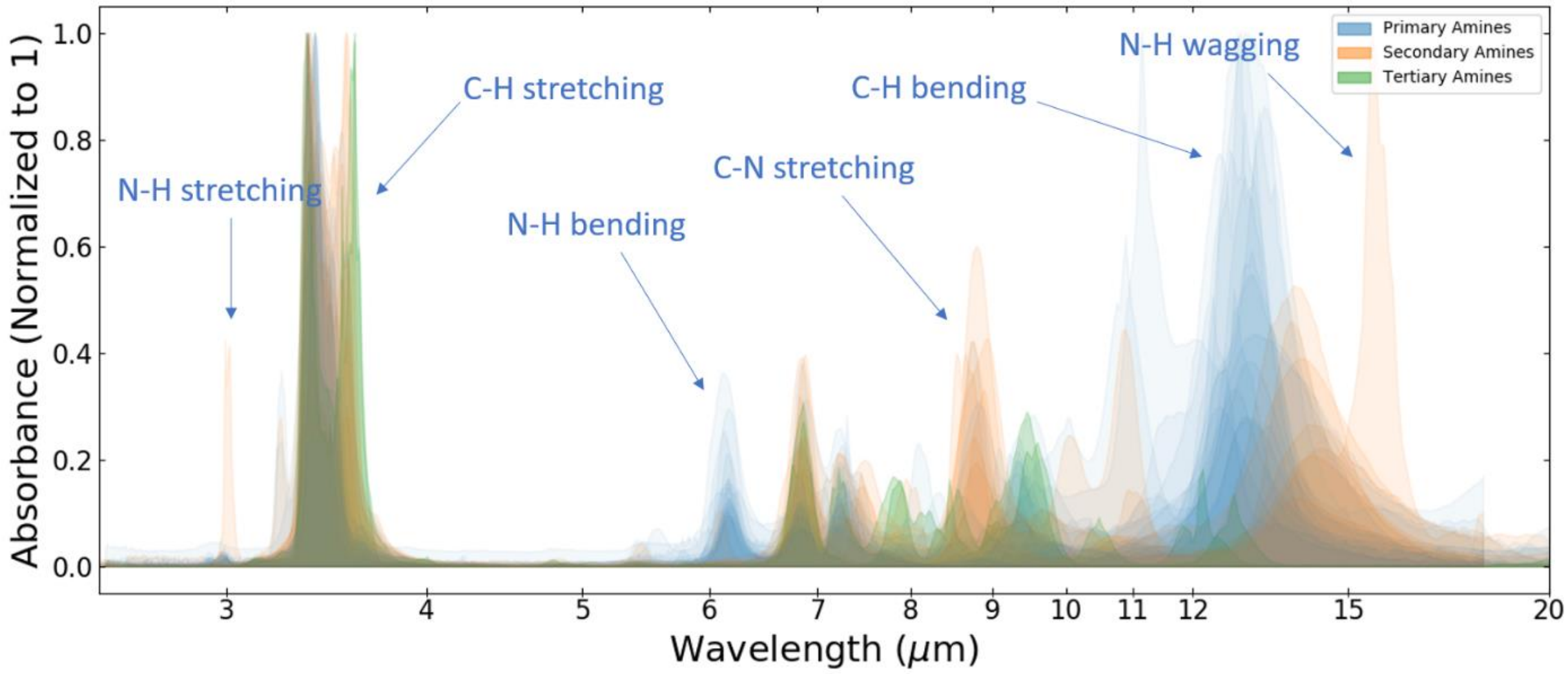

**Figure 10**. Spectral features of primary, secondary and tertiary amines in NIST (Manion et al., 2008). The x-axis shows wavelength in microns, and the y-axis shows absorbance that is normalized to 1. Amines in each group show similar IR absorbance features[7].

## 5.7 $NH_3$-Induced Hazes

On an exoplanet with an $H_2$-dominated atmosphere, $NH_3$ haze generation processes have not yet been worked out in detail. In contrast to the formation of hydrocarbon and sulfur-based hazes that have been extensively studied (Domagal-Goldman et al., 2011; Arney et al., 2017; He et al., 2020), there is very little work on nitrogen-based hazes in $H_2$-dominated atmospheres. $NH_3$, on its own, does not participate in haze formation in reduced atmospheres. Hence, we do not focus on the formation and impact of $NH_3$-induced haze in our analysis. If hydrocarbons are present in the atmosphere and form hazes (Arney et al., 2017), $NH_3$ might participate in haze formation. In the presence of hydrocarbons, UV photolysis of $NH_3$ could form complex nitrogen-containing organics. This process is similar to the formation of organophosphine hazes (Sousa-Silva et al., 2020).  However, contrary to the formation of organophosphine hazes, UV photolysis of $NH_3$ could reduce hydrocarbon hazes. The nitrogen-containing organics formed from UV photolysis of $NH_3$ are amines in nature. They are highly soluble and are susceptible to atmospheric precipitation.

It's worth noting that on Earth, $NH_3$ plays a vital role in haze formation, especially $PM_{2.5}$ (i.e., fine particulate matter with diameter <2.5 μm) (Ye et al., 2011; Wei et al., 2015; Chen et al., 2016; Liu et al., 2019). In Earth's oxidized atmosphere, one way that atmospheric $NH_3$ can induce haze formation is by reacting with other pollutants such as sulfur dioxide ($SO_2$) and nitrogen oxides ($NO_x$). These reactions form fine airborne particles such as ammonium nitrate ($NH_4NO_3$), ammonium bisulfate ($NH_4HSO_4$), and ammonium sulfate ($(NH_4)_2SO_4$). The airborne particles can then form $PM_{2.5}$ by mixing with dust and other air pollutants (Chen et al., 2016). Furthermore, it has been shown that $NH_3$ can facilitate oxidation-reduction between $NO_2$ and $SO_2$ to form atmospheric nitrous acid (HONO) (Ge et al., 2019). Under UV conditions, atmospheric HONO can undergo photolysis to yield OH radicals, further facilitating haze formation (Ge et al., 2019; Wang et al., 2020).

---

[7] Some peaks remain undiagnosed.

### 5.8 $NH_3$ as a greenhouse gas

Ammonia is a greenhouse gas, given that it is a good IR absorber. However, its greenhouse effect is negligible on Earth due to its high reactivity and short atmospheric lifetime.

For exoplanets, we show in a separate paper that an increase in atmospheric $NH_3$ concentration does not lead to a considerable increase in surface temperature (Ranjan et al. Submitted). Specifically, in an $H_2$-dominated atmosphere, the surface temperature only increases by about 10 K with 100 ppm of atmospheric $NH_3$ due to $NH_3$-$H_2$ collision-induced absorption. In contrast, for an $N_2$-dominated atmosphere, the temperature increase is roughly 40 K with 100 ppm of $NH_3$ (Ranjan et al. Submitted).

In an implausible scenario where $NH_3$ does accumulate to extremely high concentrations on exoplanets, $NH_3$ can indeed heat the surface significantly, particularly for $N_2$- and $CO_2$-dominated atmospheres. $NH_3$-induced heating on $H_2$-dominated atmospheres is minimal because $NH_3$ absorption is degenerate with $H_2$-$H_2$ collision-induced absorption. Such high concentrations of $NH_3$ might also heat the stratosphere due to $NH_3$ UV absorption.

### 5.9 $NH_3$ Detectability with Future Telescopes

We assess the detectability of $NH_3$ with the premise of using a JWST-like telescope, which has a 6.5-meter diameter primary mirror, an estimated systematic noise floor of 10 ppm, and an estimated service time of 5 years (cryogenic lifetime). As discussed in Section 4.3, for exoplanets transiting M5V stars and orbiting in the habitable zone, we find that JWST can characterize only those with an $H_2$-dominated atmosphere with reasonable observation time. However, not all terrestrial exoplanets can retain $H_2$-dominated atmospheres.

To explore the potential for detecting $NH_3$ in a non-$H_2$-dominated atmosphere, we assume observation using more capable telescopes that are currently in design or development, such as missions like The Origins Space Telescope (OST) (Battersby et al., 2018) and the 30-meter class of ground telescopes (Johns et al., 2012; Tamai and Spyromilio 2014; Skidmore et al., 2015). Since these telescopes' instrumental details may change with mission development, we generalize those details into two abstract categories: (1) a 10-meter space telescope with broad spectral coverage. (2) a 30-meter ground-based telescope constrained by Earth’s atmosphere observing windows. In both scenarios, we assume a 1 ppm noise floor. We computed the new noise estimate by scaling the JWST noise simulator Pandexo output with lower noise floor input.

For a 10-meter space telescope with a lower noise floor, characterization of $CO_2$-dominated or $N_2$-dominated atmospheres for exoplanets transiting an M5V star (which have an atmospheric transit-depth of 10 ~ 20 ppm) are possible with 20 transits per observation mode. However, for exoplanets transiting a Sun-like star, even $H_2$-dominated atmospheres are not accessible (atmospheric transit-depth of 2 ~ 3 ppm). The fundamental constraint here is $(R_{planet}/R_{star})^2$. While it is arguably possible to characterize such $H_2$-dominated atmospheres with a 30-meter ground-based telescope, the only $NH_3$ IR spectral features observable in Earth's atmosphere window are the 2.3 μm and the 10 μm features.

## 5.10 $NH_3$ in Mini-Neptunes

Super-Earths and mini-Neptunes discovered by future direct imaging programs may not be distinguishable from each other, lacking a mass and/or a size measurement. In this situation, any $NH_3$ detected should not be considered a possible biosignature gas. The reason is that $NH_3$ should exist to some level in mini-Neptune atmospheres without production by life, as $NH_3$ can be generated deep in the mini-Neptune envelope where temperature and pressure are high enough for $NH_3$'s atmospheric production. More work is needed to explore mini-Neptunes, including photochemistry, where $NH_3$ might be a marker for identifying a directly imaged planet as a mini-Neptune instead of a super-Earth.

## 6. Summary

In this paper, we examine the potential of $NH_3$ as a biosignature gas. We use various approaches, ranging from comparing Henry's law constants for different atmospheric gases to our simplified ocean-$NH_3$ interaction model to applications of our comprehensive photochemistry code and transmission spectra model.

In brief, $NH_3$ in a terrestrial planet atmosphere is generally a good biosignature gas, primarily because terrestrial planets have no significant known abiotic $NH_3$ source. $NH_3$'s high water solubility and high bio-useability likely prevent $NH_3$ from accumulating in the atmosphere to detectable levels unless life is a net source of $NH_3$ and produces enough $NH_3$ to saturate the surface sinks. Only then can $NH_3$ accumulate in the atmosphere with a reasonable surface production flux.

Specifically, for the favorable scenario of exoplanets with $H_2$-dominated atmospheres orbiting M dwarf stars (M5V), we find that a minimum of about 5 ppm column-averaged mixing ratio is needed for $NH_3$ to be detectable with JWST, considering a 10 ppm JWST systematic noise floor.

Additionally, volatile amines share $NH_3$'s weaknesses and strengths as a biosignature since volatile amines have similar solubilities and reactivities to $NH_3$. Finally, to establish $NH_3$ as a biosignature gas, we must rule out mini-Neptunes with deep atmospheres, where temperatures and pressures are high enough for $NH_3$'s atmospheric production.

## Acknowledgments

We thank William Bains for the valuable discussions. We thank Thomas Evans, Ana Glidden, and Zahra Essack for helpful discussion to establish the detection metric. We acknowledge funding from the Heising-Simons Foundation (grant number #2018-1104) and NASA (grant numbers 80NSSC19K0471 and NNX15AC86G). S.R. gratefully acknowledges support from the Simons Foundation, grant no. 495062.

# Supplementary Information

## SI.1 Simulation Parameters for the Photochemistry Model

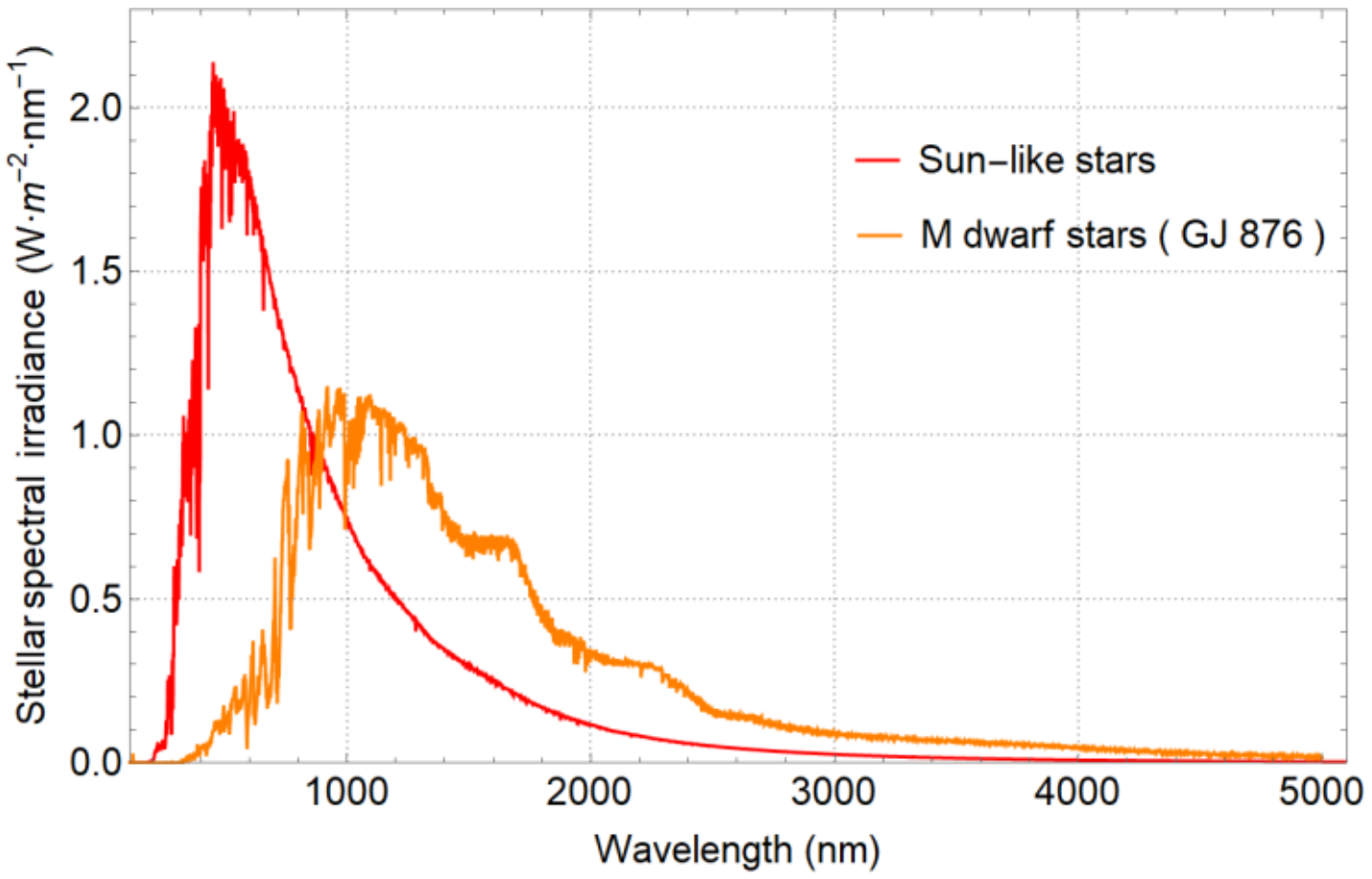

**Figure SI.1**. The synthetic stellar spectrum input of our photochemistry model (Loyd et al., 2016; France et al., 2016). The y-axis shows spectral irradiance in W·m$^{-2}$·nm$^{-1}$, and the x-axis shows wavelength in nm.

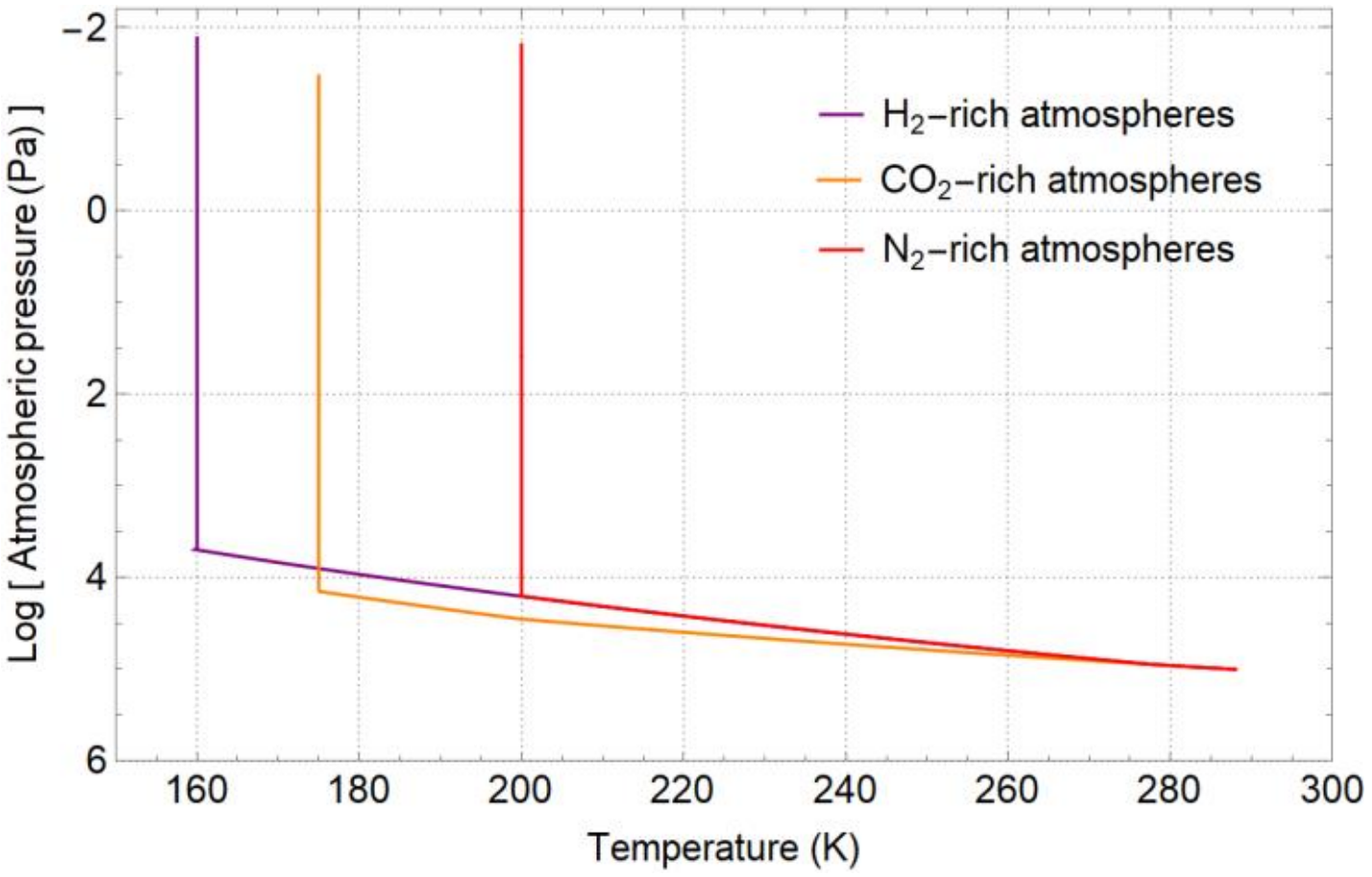

**Figure SI.2**. The temperature-pressure profiles of the simulated exoplanets with $H_2$-dominated, $CO_2$-dominated, and $N_2$-dominated atmospheres. The y-axis shows atmospheric pressure (Pa) on a log scale, and the x-axis shows temperature in Kelvin (K).

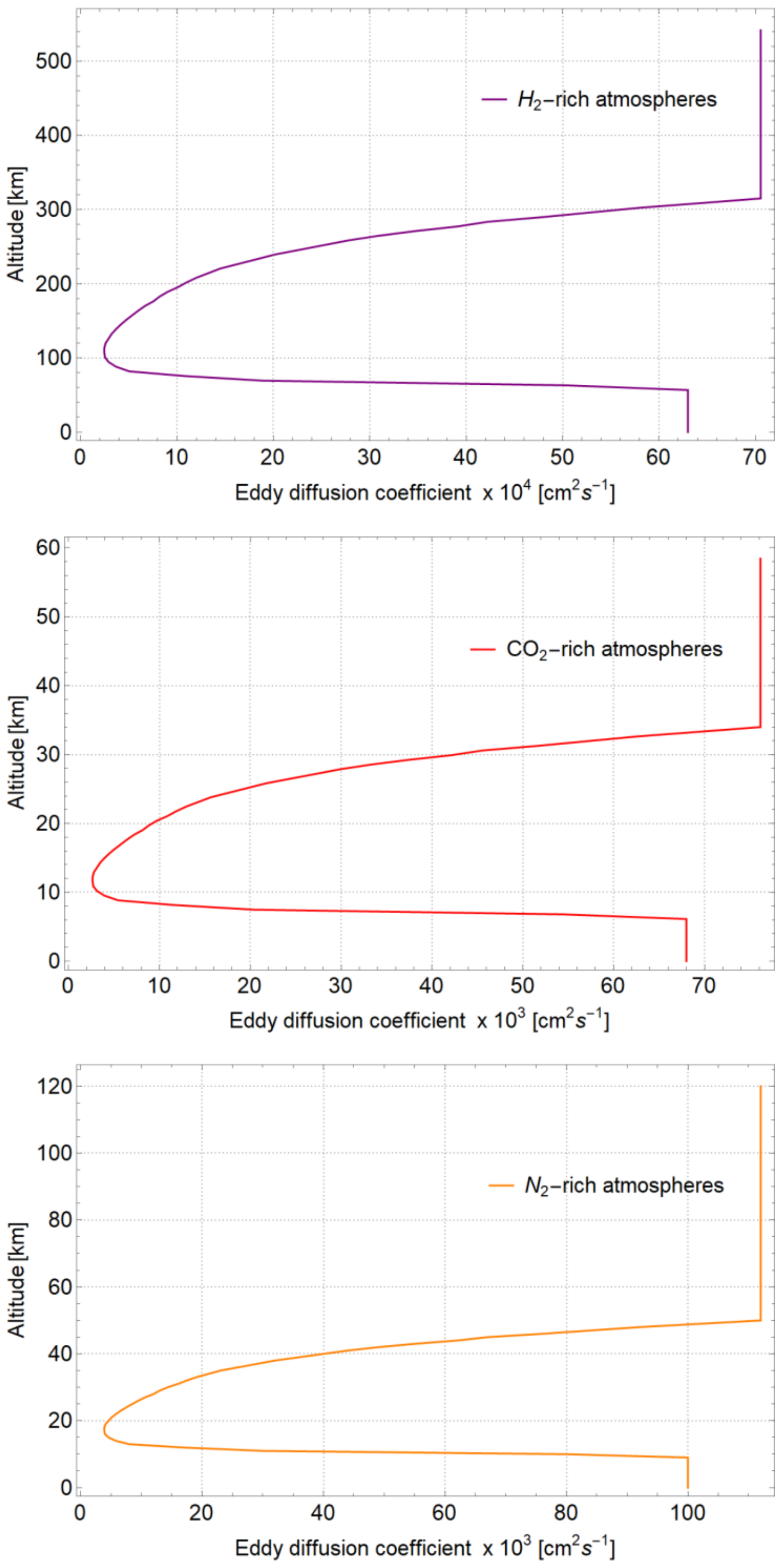

**Figure SI.3**. The eddy diffusion profiles of the simulated exoplanets with $H_2$-dominated (top panel), $CO_2$-dominated (middle panel), and $N_2$-dominated (bottom panel) atmospheres. The y-axis shows altitude in km, and the x-axis shows the eddy diffusion coefficient in $cm^2 \cdot s^{-1}$.

Our photochemistry model demarcates chemical species into 4 types: type “A” for aerosol species, type “C” for chemically inert species which are assumed not to react or transport, type “F” for species assumed to be in photochemical equilibrium (i.e., for which transport is neglected), and type "X", for species for which the full photochemical transport equation is solved. For the lower boundary condition, our model allows us to either specify a fixed surface mixing ratio (type "1") or the surface emission flux and dry deposition velocity (type "2") (Hu et al., 2012).

**Table S1-1.** Surface boundary conditions for exoplanets with $CO_2$-dominated atmospheres.

| **Name** | **Type** | **Initial Mixing Ratio** | **Upper Boundary Flux Beside Escape (Upwards) [molecule/(cm$^2$s)]** | **Lower Boundary Type** | **Dry Deposition Velocity [cm/s]** | **Lower Boundary Flux (Upwards) [molecule/(cm$^2$s)]** |
|---|---|---|---|---|---|---|
| H | X | 0 | 0 | 2 | 1 | 0 |
| $H_2$ | X | $1.60\times10^{-3}$ | 0 | 2 | 0 | $3.00\times10^{10}$ |
| O | X | 0 | 0 | 2 | 1 | 0 |
| O(1D) | X | 0 | 0 | 2 | 0 | 0 |
| $O_2$ | X | 0 | 0 | 2 | 0 | 0 |
| $O_3$ | X | 0 | 0 | 2 | 0.4 | 0 |
| OH | X | 0 | 0 | 2 | 1 | 0 |
| $HO_2$ | X | 0 | 0 | 2 | 1 | 0 |
| $H_2O$ | X | $2.00\times10^{-6}$ | 0 | 1 | 0 | 0.01 |
| $H_2O_2$ | X | 0 | 0 | 2 | 0.5 | 0 |
| $CO_2$ | X | 0.9 | 0 | 1 | 0 | 0.9 |
| CO | X | 0 | 0 | 2 | $1.00\times10^{-8}$ | 0 |
| $CH_2O$ | X | 0 | 0 | 2 | 0.1 | 0 |
| CHO | X | 0 | 0 | 2 | 0.1 | 0 |
| C | X | 0 | 0 | 2 | 0 | 0 |
| CH | X | 0 | 0 | 2 | 0 | 0 |
| $CH_2$ | X | 0 | 0 | 2 | 0 | 0 |
| $CH_{21}$ | X | 0 | 0 | 2 | 0 | 0 |
| $CH_3$ | X | 0 | 0 | 2 | 0 | 0 |
| $CH_4$ | X | 0 | 0 | 2 | 0 | $3.00\times10^{8}$ |
| $CH_3O$ | X | 0 | 0 | 2 | 0.1 | 0 |
| $CH_4O$ | X | 0 | 0 | 2 | 0.1 | 0 |
| $CHO_2$ | X | 0 | 0 | 2 | 0.1 | 0 |
| $CH_2O_2$ | X | 0 | 0 | 2 | 0.1 | 0 |
| $CH_3O_2$ | X | 0 | 0 | 2 | 0 | 0 |
| $CH_4O_2$ | X | 0 | 0 | 2 | 0.1 | 0 |
| C2 | X | 0 | 0 | 2 | 0 | 0 |
| $C_2H$ | X | 0 | 0 | 2 | 0 | 0 |
| $C_2H_2$ | X | 0 | 0 | 2 | 0 | 0 |
| $C_2H_3$ | X | 0 | 0 | 2 | 0 | 0 |
| $C_2H_4$ | X | 0 | 0 | 2 | 0 | 0 |
| $C_2H_5$ | X | 0 | 0 | 2 | 0 | 0 |
| $C_2H_6$ | X | 0 | 0 | 2 | $1.00\times10^{-5}$ | 0 |
| $C_2HO$ | X | 0 | 0 | 2 | 0 | 0 |
| $C_2H_2O$ | X | 0 | 0 | 2 | 0.1 | 0 |
| $C_2H_3O$ | X | 0 | 0 | 2 | 0.1 | 0 |
| $C_2H_4O$ | X | 0 | 0 | 2 | 0.1 | 0 |
| $C_2H_5O$ | X | 0 | 0 | 2 | 0.1 | 0 |
| S | X | 0 | 0 | 2 | 0 | 0 |
| $S_2$ | X | 0 | 0 | 2 | 0 | 0 |
| $S_3$ | X | 0 | 0 | 2 | 0 | 0 |
| $S_4$ | X | 0 | 0 | 2 | 0 | 0 |
| SO | X | 0 | 0 | 2 | 0 | 0 |
| $SO_2$ | X | 0 | 0 | 2 | 1 | $3.00\times10^9$ |
| $SO_{21}$ | X | 0 | 0 | 2 | 0 | 0 |
| $SO_{23}$ | X | 0 | 0 | 2 | 0 | 0 |
| $SO_3$ | X | 0 | 0 | 2 | 1 | 0 |
| $H_2S$ | X | 0 | 0 | 2 | 0.015 | $3.00\times10^8$ |
| HS | X | 0 | 0 | 2 | 0 | 0 |
| HSO | X | 0 | 0 | 2 | 0 | 0 |
| $HSO_2$ | X | 0 | 0 | 2 | 0 | 0 |
| $HSO_3$ | X | 0 | 0 | 2 | 0.1 | 0 |
| $H_2SO_4$ | X | 0 | 0 | 2 | 1 | 0 |
| $H_2SO_4A$ | A | 0 | 0 | 2 | 0.2 | 0 |
| $S_8$ | X | 0 | 0 | 2 | 0 | 0 |
| $S_8A$ | A | 0 | 0 | 2 | 0.2 | 0 |
| $N_2$ | C | 0.1 | 0 | 1 | 0 | 0.1 |
| N | X | 0 | 0 | 2 | 0 | 0 |
| $NH_3$ | X | 0 | 0 | 2 | 1 | $1.0\times10^9$ |
| $NH_2$ | X | 0 | 0 | 2 | 0 | 0 |
| NH | X | 0 | 0 | 2 | 0 | 0 |
| $N_2O$ | X | 0 | 0 | 2 | 0 | 0 |
| NO | X | 0 | 0 | 2 | 0.02 | 0 |
| $NO_2$ | X | 0 | 0 | 2 | 0.02 | 0 |
| $NO_3$ | X | 0 | 0 | 2 | 1 | 0 |
| $N_2O_5$ | X | 0 | 0 | 2 | 4 | 0 |
| HNO | X | 0 | 0 | 2 | 0 | 0 |
| $HNO_2$ | X | 0 | 0 | 2 | 0.5 | 0 |
| $HNO_3$ | X | 0 | 0 | 2 | 4 | 0 |
| $HNO_4$ | X | 0 | 0 | 2 | 4 | 0 |
| HCN | X | 0 | 0 | 2 | 0.01 | 0 |
| CN | X | 0 | 0 | 2 | 0.01 | 0 |
| CNO | X | 0 | 0 | 2 | 0 | 0 |
| HCNO | X | 0 | 0 | 2 | 0 | 0 |
| $CH_3NO_2$ | X | 0 | 0 | 2 | 0.01 | 0 |
| $CH_3NO_3$ | X | 0 | 0 | 2 | 0.01 | 0 |
| $CH_5N$ | X | 0 | 0 | 2 | 0 | 0 |
| $C_2H_2N$ | X | 0 | 0 | 2 | 0 | 0 |
| $C_2H_5N$ | X | 0 | 0 | 2 | 0 | 0 |
| $N_2H_2$ | X | 0 | 0 | 2 | 0 | 0 |
| $N_2H_3$ | X | 0 | 0 | 2 | 0 | 0 |
| $N_2H_4$ | X | 0 | 0 | 2 | 0 | 0 |
| OCS | X | 0 | 0 | 2 | 0.01 | 0 |
| CS | X | 0 | 0 | 2 | 0.01 | 0 |
| $CH_3S$ | X | 0 | 0 | 2 | 0.01 | 0 |
| $CH_4S$ | X | 0 | 0 | 2 | 0.01 | 0 |

**Table S1-2.** Surface boundary conditions for exoplanets with $H_2$-dominated atmospheres.

| Name | Type | Initial Mixing Ratio | Upper Boundary Flux Beside Escape (Upwards) [molecule/(cm$^2$s)] | Lower Boundary Type | Dry Deposition Velocity [cm/s] | Lower Boundary Flux (Upwards) [molecule/(cm$^2$s)] |
|---|---|---|---|---|---|---|
| H | X | 0 | 0 | 2 | 1 | 0 |
| $H_2$ | C | 0.9 | 0 | 1 | 0 | 0.9 |
| O | X | 0 | 0 | 2 | 1 | 0 |
| O(1D) | X | 0 | 0 | 2 | 0 | 0 |
| $O_2$ | X | 0 | 0 | 2 | 0 | 0 |
| $O_3$ | X | 0 | 0 | 2 | 0 | 0 |
| OH | X | 0 | 0 | 2 | 1 | 0 |
| $HO_2$ | X | 0 | 0 | 2 | 1 | 0 |
| $H_2O$ | X | $2.00\times10^{-6}$ | 0 | 1 | 0 | $1.00\times10^{-2}$ |
| $H_2O_2$ | X | 0 | 0 | 2 | 0.5 | 0 |
| $CO_2$ | X | 0 | 0 | 2 | $1.00\times10^{-4}$ | $3.00\times10^{11}$ |
| CO | X | 0 | 0 | 2 | $1.00\times10^{-8}$ | 0 |
| $CH_2O$ | X | 0 | 0 | 2 | 0.1 | 0 |
| CHO | X | 0 | 0 | 2 | 0.1 | 0 |
| C | X | 0 | 0 | 2 | 0 | 0 |
| CH | X | 0 | 0 | 2 | 0 | 0 |
| $CH_2$ | X | 0 | 0 | 2 | 0 | 0 |
| $CH_{21}$ | X | 0 | 0 | 2 | 0 | 0 |
| $CH_3$ | X | 0 | 0 | 2 | 0 | 0 |
| $CH_4$ | X | 0 | 0 | 2 | 0 | $3.00\times10^{8}$ |
| $CH_3O$ | X | 0 | 0 | 2 | 0.1 | 0 |
| $CH_4O$ | X | 0 | 0 | 2 | 0.1 | 0 |
| $CH_2O_2$ | X | 0 | 0 | 2 | 0.1 | 0 |
| $CH_3O_2$ | X | 0 | 0 | 2 | 0 | 0 |
| $CH_4O_2$ | X | 0 | 0 | 2 | 0.1 | 0 |
| $C_2$ | X | 0 | 0 | 2 | 0 | 0 |
| $C_2H$ | X | 0 | 0 | 2 | 0 | 0 |
| $C_2H_2$ | X | 0 | 0 | 2 | 0 | 0 |
| $C_2H_3$ | X | 0 | 0 | 2 | 0 | 0 |
| $C_2H_4$ | X | 0 | 0 | 2 | 0 | 0 |
| $C_2H_5$ | X | 0 | 0 | 2 | 0 | 0 |
| $C_2H_6$ | X | 0 | 0 | 2 | $1.00\times10^{-5}$ | 0 |
| $C_2HO$ | X | 0 | 0 | 2 | 0 | 0 |
| $C_2H_2O$ | X | 0 | 0 | 2 | 0.1 | 0 |
| $C_2H_3O$ | X | 0 | 0 | 2 | 0.1 | 0 |
| $C_2H_4O$ | X | 0 | 0 | 2 | 0.1 | 0 |
| $C_2H_5O$ | X | 0 | 0 | 2 | 0.1 | 0 |
| $N_2$ | C | 0.1 | 0 | 1 | 0 | 0.1 |
| S | X | 0 | 0 | 2 | 0 | 0 |
| $S_2$ | X | 0 | 0 | 2 | 0 | 0 |
| $S_3$ | X | 0 | 0 | 2 | 0 | 0 |
| $S_4$ | X | 0 | 0 | 2 | 0 | 0 |
| SO | X | 0 | 0 | 2 | 0 | 0 |
| $SO_2$ | X | 0 | 0 | 2 | 1 | $3.00\times10^{9}$ |
| $SO_{21}$ | X | 0 | 0 | 2 | 0 | 0 |
| $SO_{23}$ | X | 0 | 0 | 2 | 0 | 0 |
| $SO_3$ | X | 0 | 0 | 2 | 0 | 0 |
| $H_2S$ | X | 0 | 0 | 2 | 0.015 | $3.00\times10^{8}$ |
| HS | X | 0 | 0 | 2 | 0 | 0 |
| HSO | X | 0 | 0 | 2 | 0 | 0 |
| $HSO_2$ | X | 0 | 0 | 2 | 0 | 0 |
| $HSO_3$ | X | 0 | 0 | 2 | 0 | 0 |
| $H_2SO_4$ | X | 0 | 0 | 2 | 1 | 0 |
| $H_2SO_4A$ | A | 0 | 0 | 2 | 0.2 | 0 |
| $S_8$ | X | 0 | 0 | 2 | 0 | 0 |
| $S_8A$ | A | 0 | 0 | 2 | 0.2 | 0 |
| $CHO_2$ | X | 0 | 0 | 2 | 0.1 | 0 |
| N | X | 0 | 0 | 2 | 0 | 0 |
| $NH_3$ | X | 0 | 0 | 2 | 1 | $1.0\times10^{10}$ |
| $NH_2$ | X | 0 | 0 | 2 | 0 | 0 |
| NH | X | 0 | 0 | 2 | 0 | 0 |
| $N_2O$ | X | 0 | 0 | 2 | 0 | 0 |
| NO | X | 0 | 0 | 2 | 0.02 | 0 |
| $NO_2$ | X | 0 | 0 | 2 | 0.02 | 0 |
| $NO_3$ | X | 0 | 0 | 2 | 1 | 0 |
| $N_2O_5$ | X | 0 | 0 | 2 | 4 | 0 |
| HNO | X | 0 | 0 | 2 | 0 | 0 |
| $HNO_2$ | X | 0 | 0 | 2 | 0.5 | 0 |
| $HNO_3$ | X | 0 | 0 | 2 | 4 | 0 |
| $HNO_4$ | X | 0 | 0 | 2 | 4 | 0 |
| HCN | X | 0 | 0 | 2 | 0.01 | 0 |
| CN | X | 0 | 0 | 2 | 0.01 | 0 |
| CNO | X | 0 | 0 | 2 | 0 | 0 |
| HCNO | X | 0 | 0 | 2 | 0 | 0 |
| $CH_3NO_2$ | X | 0 | 0 | 2 | 0.01 | 0 |
| $CH_3NO_3$ | X | 0 | 0 | 2 | 0.01 | 0 |
| $CH_5N$ | X | 0 | 0 | 2 | 0 | 0 |
| $C_2H_2N$ | X | 0 | 0 | 2 | 0 | 0 |
| $C_2H_5N$ | X | 0 | 0 | 2 | 0 | 0 |
| $N_2H_2$ | X | 0 | 0 | 2 | 0 | 0 |
| $N_2H_3$ | X | 0 | 0 | 2 | 0 | 0 |
| $N_2H_4$ | X | 0 | 0 | 2 | 0 | 0 |
| OCS | X | 0 | 0 | 2 | 0.01 | 0 |
| CS | X | 0 | 0 | 2 | 0.01 | 0 |
| $CH_3S$ | X | 0 | 0 | 2 | 0.01 | 0 |
| $CH_4S$ | X | 0 | 0 | 2 | 0.01 | 0 |

**Table S1-3.** Surface boundary conditions for exoplanets with $N_2$-dominated atmospheres.

| Name | Type | Initial Mixing Ratio | Upper Boundary Flux Beside Escape (Upwards) [molecule/(cm$^2$s)] | Lower Boundary Type | Dry Deposition Velocity [cm/s] | Lower Boundary Flux (Upwards) [molecule/(cm$^2$s)] |
|---|---|---|---|---|---|---|
| H | X | 0 | 0 | 2 | 1 | 0 |
| $H_2$ | X | 0 | 0 | 2 | 0 | $3.00\times10^{10}$ |
| O | X | 0 | 0 | 2 | 1 | 0 |
| O(1D) | X | 0 | 0 | 2 | 0 | 0 |
| $O_2$ | X | 0 | 0 | 2 | 0 | 0 |
| $O_3$ | X | 0 | 0 | 2 | 0.4 | 0 |
| OH | X | 0 | 0 | 2 | 1 | 0 |
| $HO_2$ | X | 0 | 0 | 2 | 1 | 0 |
| $H_2O$ | X | $2.00\times10^{-6}$ | 0 | 1 | 0 | $1.00\times10^{-2}$ |
| $H_2O_2$ | X | 0 | 0 | 2 | 0.5 | 0 |
| $CO_2$ | X | 0 | 0 | 2 | $1.00\times10^{-4}$ | $3.00\times10^{11}$ |
| CO | X | 0 | 0 | 2 | $1.00\times10^{-8}$ | 0 |
| $CH_2O$ | X | 0 | 0 | 2 | 0.1 | 0 |
| CHO | X | 0 | 0 | 2 | 0.1 | 0 |
| C | X | 0 | 0 | 2 | 0 | 0 |
| CH | X | 0 | 0 | 2 | 0 | 0 |
| $CH_2$ | X | 0 | 0 | 2 | 0 | 0 |
| $CH_{21}$ | X | 0 | 0 | 2 | 0 | 0 |
| $CH_3$ | X | 0 | 0 | 2 | 0 | 0 |
| $CH_4$ | X | 0 | 0 | 2 | 0 | $3.00\times10^{8}$ |
| $CH_3O$ | X | 0 | 0 | 2 | 0.1 | 0 |
| $CH_4O$ | X | 0 | 0 | 2 | 0.1 | 0 |
| $CHO_2$ | X | 0 | 0 | 2 | 0.1 | 0 |
| $CH_2O_2$ | X | 0 | 0 | 2 | 0.1 | 0 |
| $CH_3O_2$ | X | 0 | 0 | 2 | 0.1 | 0 |
| $CH_4O_2$ | X | 0 | 0 | 2 | 0.1 | 0 |
| C2 | X | 0 | 0 | 2 | 0 | 0 |
| $C_2H$ | X | 0 | 0 | 2 | 0 | 0 |
| $C_2H_2$ | X | 0 | 0 | 2 | 0 | 0 |
| $C_2H_3$ | X | 0 | 0 | 2 | 0 | 0 |
| $C_2H_4$ | X | 0 | 0 | 2 | 0 | 0 |
| $C_2H_5$ | X | 0 | 0 | 2 | 0 | 0 |
| $C_2H_6$ | X | 0 | 0 | 2 | $1.00\times10^{-5}$ | 0 |
| $C_2HO$ | X | 0 | 0 | 2 | 0.1 | 0 |
| $C_2H_2O$ | X | 0 | 0 | 2 | 0.1 | 0 |
| $C_2H_3O$ | X | 0 | 0 | 2 | 0.1 | 0 |
| $C_2H_4O$ | X | 0 | 0 | 2 | 0.1 | 0 |
| $C_2H_5O$ | X | 0 | 0 | 2 | 0.1 | 0 |
| $N_2$ | C | 1 | 0 | 1 | 0 | 1 |
| S | X | 0 | 0 | 2 | 0 | 0 |
| $S_2$ | X | 0 | 0 | 2 | 0 | 0 |
| $S_3$ | X | 0 | 0 | 2 | 0 | 0 |
| $S_4$ | X | 0 | 0 | 2 | 0 | 0 |
| SO | X | 0 | 0 | 2 | 0 | 0 |
| $SO_2$ | X | 0 | 0 | 2 | 1 | $3.00\times10^{9}$ |
| $SO_{21}$ | X | 0 | 0 | 2 | 0 | 0 |
| $SO_{23}$ | X | 0 | 0 | 2 | 0 | 0 |
| $SO_3$ | X | 0 | 0 | 2 | 1 | 0 |
| $H_2S$ | X | 0 | 0 | 2 | 0.015 | $3.00\times10^{8}$ |
| HS | X | 0 | 0 | 2 | 0 | 0 |
| HSO | X | 0 | 0 | 2 | 0 | 0 |
| $HSO_2$ | X | 0 | 0 | 2 | 0 | 0 |
| $HSO_3$ | X | 0 | 0 | 2 | 0.1 | 0 |
| $H_2SO_4$ | X | 0 | 0 | 2 | 1 | 0 |
| $H_2SO_4A$ | A | 0 | 0 | 2 | 0.2 | 0 |
| $S_8$ | X | 0 | 0 | 2 | 0 | 0 |
| $S_8A$ | A | 0 | 0 | 2 | 0.2 | 0 |
| N | X | 0 | 0 | 2 | 0 | 0 |
| $NH_3$ | X | 0 | 0 | 2 | 1 | $1.0\times10^{10}$ |
| $NH_2$ | X | 0 | 0 | 2 | 0 | 0 |
| NH | X | 0 | 0 | 2 | 0 | 0 |
| $N_2O$ | X | 0 | 0 | 2 | 0 | 0 |
| NO | X | 0 | 0 | 2 | 0.02 | 0 |
| $NO_2$ | X | 0 | 0 | 2 | 0.02 | 0 |
| $NO_3$ | X | 0 | 0 | 2 | 1 | 0 |
| $N_2O_5$ | X | 0 | 0 | 2 | 4 | 0 |
| HNO | X | 0 | 0 | 2 | 0 | 0 |
| $HNO_2$ | X | 0 | 0 | 2 | 0.5 | 0 |
| $HNO_3$ | X | 0 | 0 | 2 | 4 | 0 |
| $HNO_4$ | X | 0 | 0 | 2 | 4 | 0 |
| HCN | X | 0 | 0 | 2 | 0.01 | 0 |
| CN | X | 0 | 0 | 2 | 0.01 | 0 |
| CNO | X | 0 | 0 | 2 | 0 | 0 |
| HCNO | X | 0 | 0 | 2 | 0 | 0 |
| $CH_3NO_2$ | X | 0 | 0 | 2 | 0.01 | 0 |
| $CH_3NO_3$ | X | 0 | 0 | 2 | 0.01 | 0 |
| $CH_5N$ | X | 0 | 0 | 2 | 0 | 0 |
| $C_2H_2N$ | X | 0 | 0 | 2 | 0 | 0 |
| $C_2H_5N$ | X | 0 | 0 | 2 | 0 | 0 |
| $N_2H_2$ | X | 0 | 0 | 2 | 0 | 0 |
| $N_2H_3$ | X | 0 | 0 | 2 | 0 | 0 |
| $N_2H_4$ | X | 0 | 0 | 2 | 0 | 0 |
| OCS | X | 0 | 0 | 2 | 0.01 | 0 |
| CS | X | 0 | 0 | 2 | 0.01 | 0 |
| $CH_3S$ | X | 0 | 0 | 2 | 0.01 | 0 |
| $CH_4S$ | X | 0 | 0 | 2 | 0.01 | 0 |

## SI.2 Dissociation Constants of Representative Amines at 25°C

**Table S1-1.** The $pK_a$ values of some representative amines at 25ºC (David R. Lide et al., 2005).

| Chemical species | pKa values |
|---|---|
| Ammonia $NH_3$ | 9.25 |
| Methylamine $CH_5N$ | 10.66 |
| Ethylamine $CH_7N$ | 10.65 |
| Propylamine $C_3H_9N$ | 10.54 |
| Isopropylamine $C_3H_9N$ | 10.63 |
| Butylamine $C_4H_{11}N$ | 10.60 |
| sec-Butylamine $C_4H_{11}N$ | 10.56 |
| tert-Butylamine $C_4H_{11}N$ | 10.68 |
| Dimethylamine $C_2H_7N$ | 10.73 |
| Diethylamine $C_4H_{11}N$ | 10.84 |
| Diisopropylamine $C_6H_{15}N$ | 11.05 |
| Trimethylamine $C_3H_9N$ | 9.80 |
| Triethylamine $C_6H_{15}N$ | 10.75 |
| Diethylmethylamine $C_5H_{13}N$ | 10.35 |

## SI.3 Reaction Rates Between Life Produced Volatile Amines and O, OH Radicals

**Table S3-1**. Reactions between life produced volatile amines and OH radicals. Reaction rates are at 298K. Second-order reactions have units of [$cm^3$ $molecule^{-1}$ $s^{-1}$] (Reference: NIST).

| Chemical | Reaction with OH radicals | Reaction rate | Reaction Order |
|---|---|---|---|
| Ammonia ($NH_3$) | $NH_3 + \cdot OH \rightarrow \cdot NH_2 + H_2O$ | $1.60 \times 10^{-13}$ | 2 |
| Ethanamine ($C_2H_5NH_2$) | $C_2H_5NH_2 + \cdot OH \rightarrow C_2H_5NH + H_2O$ | $2.77 \times 10^{-11}$ | 2 |
| 2-methylpropan-2-amine (tert-$C_4H_9NH_2$) | tert-$C_4H_9NH_2$ + ·OH → $C_4H_9NH$ + $H_2O$ | $1.20 \times 10^{-11}$ | 2 |
| Trimethylamine ($C_3H_9N$) | $C_3H_9N$ + ·OH → $C_3H_8N$ + $H_2O$ | $6.11 \times 10^{-11}$ | 2 |

**Table S3-2**. Reactions between life produced volatile amines and O radicals. Reaction rates are at 298K. Second-order reactions have units of [$cm^3$ $molecule^{-1}$ $s^{-1}$] (Reference: NIST).

| Chemical | Reaction with O radicals | Reaction rate | Reaction Order |
|---|---|---|---|
| Ammonia ($NH_3$) | $NH_3$ + ·O → ·OH + ·$NH_2$ | $4.37 \times 10^{-17}$ | 2 |
| Methylamine ($CH_3NH_2$) | $CH_3NH_2$ + ·O → ·OH + $CH_3NH$ | $5.56 \times 10^{-13}$ | 2 |
| Ethylamine ($C_2H_5NH_2$) | $C_2H_5NH_2$ + ·O → ·OH + $C_2H_5NH$ | $3.87 \times 10^{-12}$ | 2 |
| Dimethylamine ($(CH_3)_2NH$) | $(CH_3)_2NH$ + ·O → ·OH + $(CH_3)_2N$· | $6.00 \times 10^{-12}$ | 2 |
| Trimethylamine ($C_3H_9N$) | $C_3H_9N$ + ·O →·OH + $C_3H_8N$ | $7.54 \times 10^{-12}$ | 2 |

## SI.4 IR Spectra of Primary, Secondary and Tertiary Amines

We list the IUPAC names of the 15 primary amines, 9 secondary amines, and 2 tertiary amines with absorbance data in NIST (Manion et al., 2008) in Table S4-1.

**Table S4-1.** Primary (15), Secondary (9) and Tertiary (2) Amines in NIST

| | |
|---|---|
| Primary amines | 2-methylpropan-2-amine, 2-methylbutan-2-amine, propan-2-amine, pentan-2-amine, methylamine, 2-methylprop-2-en-1-amine, 2-methylbutan-1-amine, ethanamine, prop-2-en-1-amine, propan-1-amine, butan-1-amine, pentan-1-amine, butane-1,4-diamine, (3-aminopropyl)(methyl)amine, (2-aminoethyl)(ethyl)amine |
| Secondary amines | diethylamine, ethyl(prop-2-en-1-yl)amine, [(1R)-1-(naphthalen-1-yl)ethyl]({3-[3-(trifluoromethyl)phenyl]propyl})amine, methyl(prop-2-yn-1-yl)amine, methyl(2-methylpropyl)amine, methyl(prop-2-en-1-yl)amine, methyl[2-(methylamino)ethyl]amine, (3-aminopropyl)(methyl)amine, (2-aminoethyl)(ethyl)amine |
| Tertiary amines | trimethylamine, ethyldimethylamine |

Here we show the overlaid IR spectra of the primary, secondary and tertiary amines.

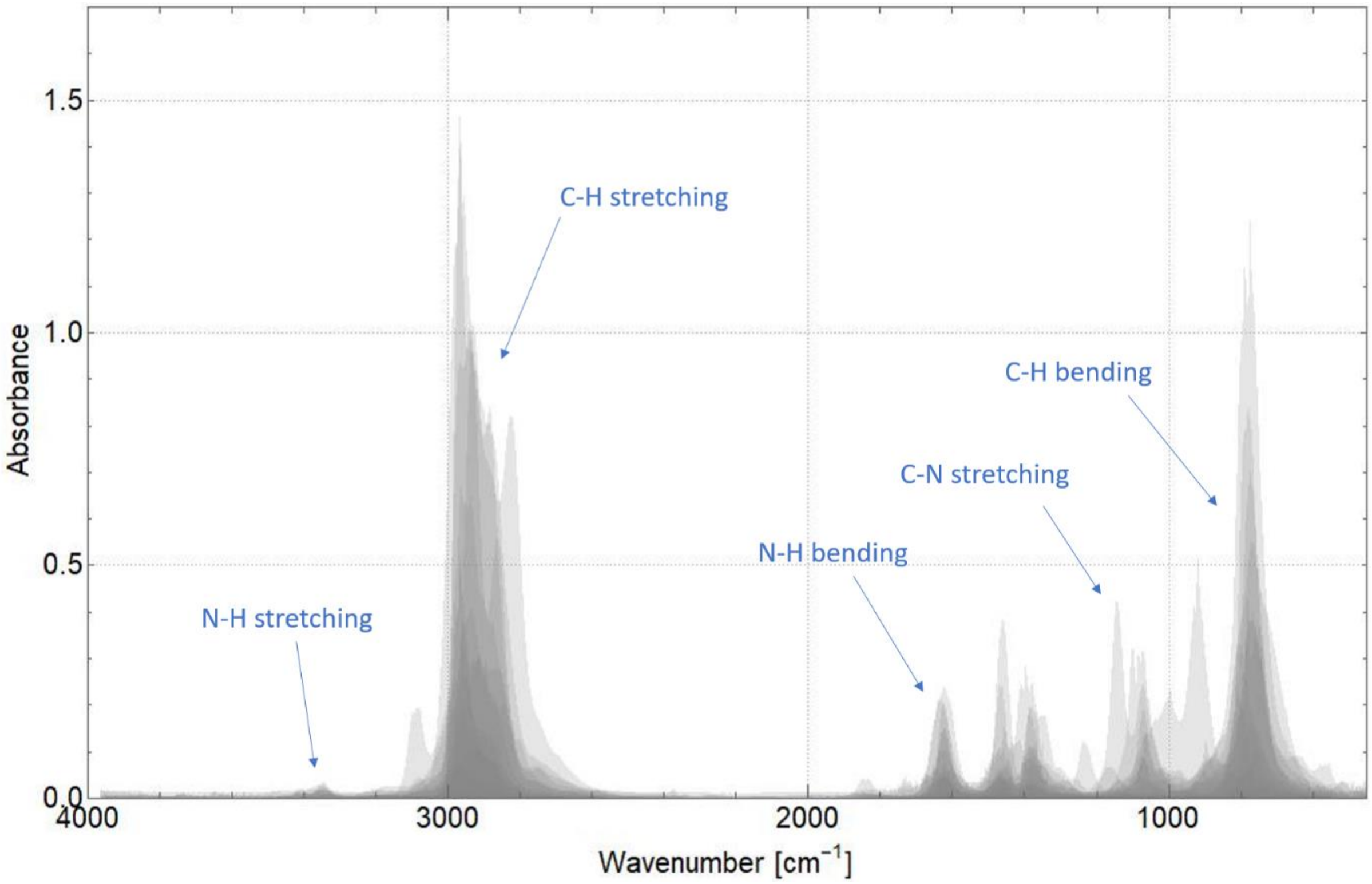

**Figure SI.4**. IR absorbance spectra of primary amines. Included are the fifteen are in the NIST database (Manion et al., 2008). The x-axis is the wavenumber, and the y axis is absorbance. Although only two of the amines contain nitrogen (N-H bending and C-N stretching), there are several distinctive peaks.

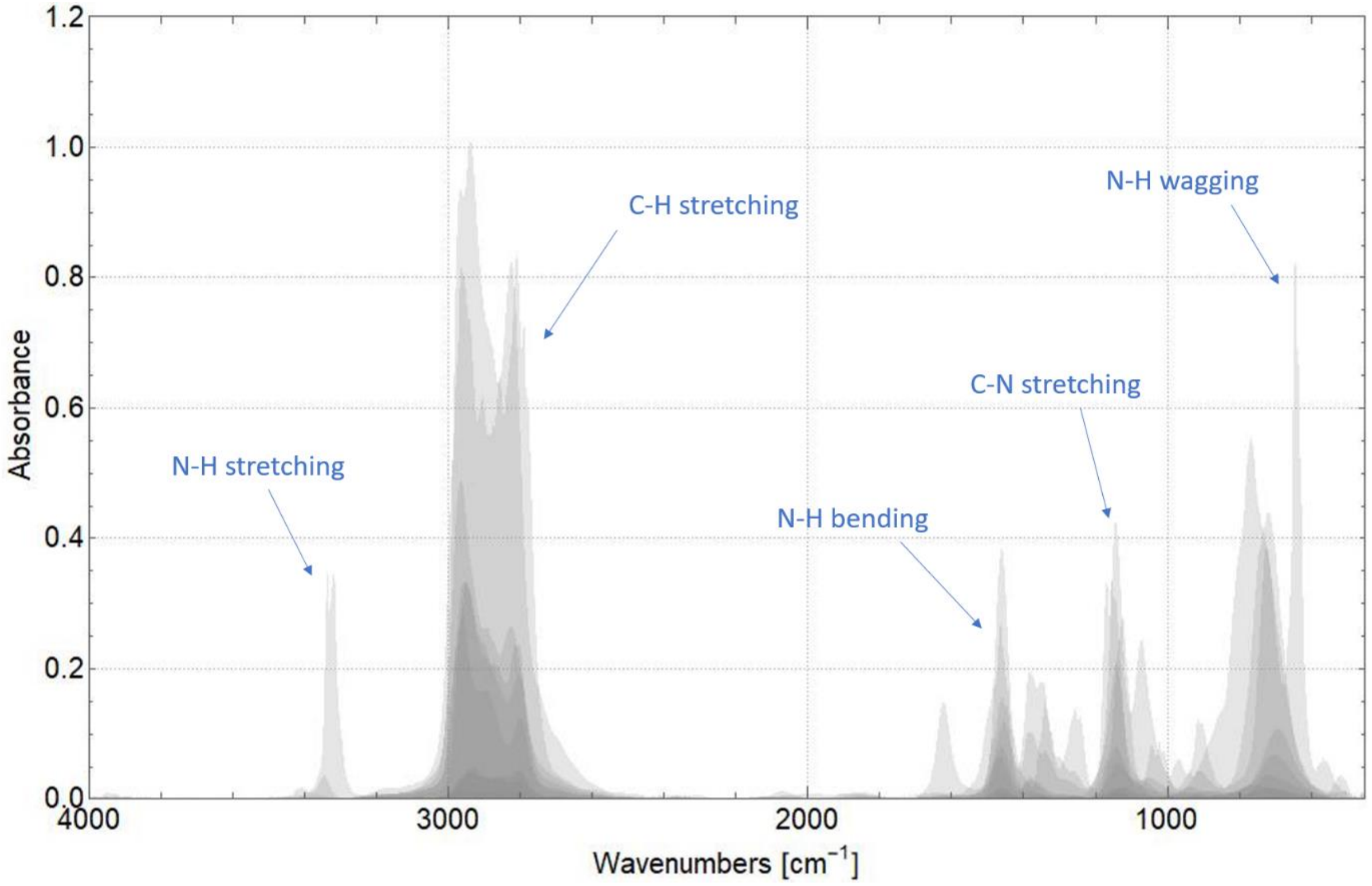

**Figure SI.5**. IR absorbance spectra of secondary amines. Nine species are present in the NIST database (Manion et al., 2008). The x-axis is the wavenumber, and the y axis is absorbance. Although only two of the amines contain nitrogen (N-H bending and C-N stretching), there are several distinctive peaks. The secondary amines have an N-H stretching feature not present in primary or tertiary amine spectra.

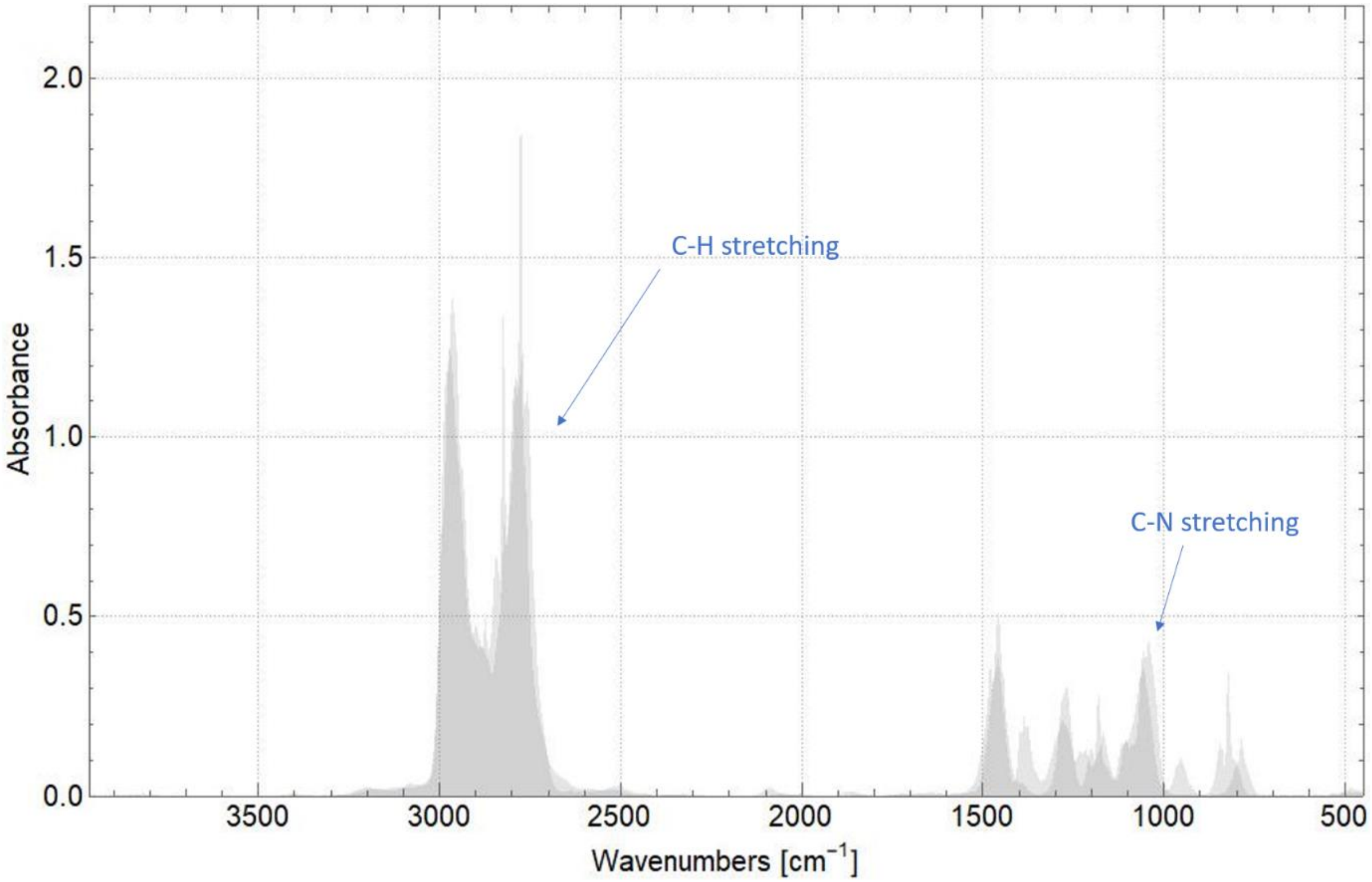

**Figure SI.6**. IR absorbance spectra of tertiary amines. Only two are present in the NIST database (Manion et al., 2008). The x-axis is the wavenumber, and the y-axis is absorbance. Tertiary amines lack N-H bonds, so they lack the N-H stretch feature found in primary and secondary amines.

## SI.5 $NH_3$ Removal Rate in the Photochemistry Model

**Table S5-1.** The top three photochemical removal mechanisms for $NH_3$ on exoplanets with $H_2$-dominated, $CO_2$-dominated, and $N_2$-dominated atmospheres orbiting M dwarf stars (M5V).

| | Atmospheric scenarios | The top three loss reactions | Loss rate [molecule $cm^{-2}$ $s^{-1}$] |
|---|---|---|---|
| With $NH_3$ deposition | $H_2$-dominated | $NH_3 \rightarrow NH_2 + H$ | $1.54 \times 10^{10}$ |
| | | $OH + NH_3 \rightarrow H_2O + NH_2$ | $1.27 \times 10^{4}$ |
| | | $H + NH_3 \rightarrow H_2 + NH_2$ | $4.99 \times 10^{2}$ |
| | $CO_2$-dominated | $NH_3 \rightarrow NH_2 + H$ | $8.35 \times 10^{8}$ |
| | | $O(1D) + NH_3 \rightarrow OH + NH_2$ | $3.30 \times 10^{6}$ |
| | | $NH_3 \rightarrow NH + H_2$ | $1.34 \times 10^{6}$ |
| | $N_2$-dominated | $NH_3 \rightarrow NH_2 + H$ | $9.01 \times 10^{10}$ |
| | | $OH + NH_3 \rightarrow H_2O + NH_2$ | $1.85 \times 10^{8}$ |
| | | $NH_3 \rightarrow NH + H_2$ | $5.60 \times 10^{5}$ |
| Without $NH_3$ deposition | $H_2$-dominated | $NH_3 \rightarrow NH_2 + H$ | $2.00 \times 10^{10}$ |
| | | $OH + NH_3 \rightarrow H_2O + NH_2$ | $2.78 \times 10^{4}$ |

| | | |
|---|---|---|
| | $NH_3 + CH \rightarrow HCN + H_2 + H$ | $1.34 \times 10^3$ |
| $CO_2$-dominated | $NH_3 \rightarrow NH_2 + H$ | $8.79 \times 10^8$ |
| | $O(1D) + NH_3 \rightarrow OH + NH_2$ | $3.54 \times 10^6$ |
| | $NH_3 \rightarrow NH + H_2$ | $1.44 \times 10^6$ |
| $N_2$-dominated | $NH_3 \rightarrow NH_2 + H$ | $9.64 \times 10^{10}$ |
| | $OH + NH_3 \rightarrow H_2O + NH_2$ | $1.96 \times 10^8$ |
| | $NH_3 \rightarrow NH + H_2$ | $1.08 \times 10^6$ |

## SI.6 Photochemistry Model Sensitivity Tests

To test the robustness of our photochemistry results, we decide to run various sensitivity tests. We choose the $H_2$-dominated atmosphere as our example here since it is the most favorable scenario for detecting $NH_3$. Here we present our results.

I. The presence of a cold trap: Our results are not sensitive to the presence of a cold trap. To first order, the presence of a cold trap does not affect $NH_3$ rainout or dry deposition. Instead, whether a cold trap is present only influences the $NH_3$ photochemical production and loss rate. Here, we simulate an atmosphere with a reduced cold trap (compared to the existing cold trap we set for the $H_2$-dominated atmosphere) (see Figure SI.7). In this case, the planet has a hot stratosphere due to UV-absorbers present in the upper atmosphere. We choose the $H_2$-dominated atmosphere as our example here since it is the most favorable scenario for detecting $NH_3$. To simulate an exoplanet with a reduced cold trap, we set the temperature above tropopause to 220K (see Figure SI.7)

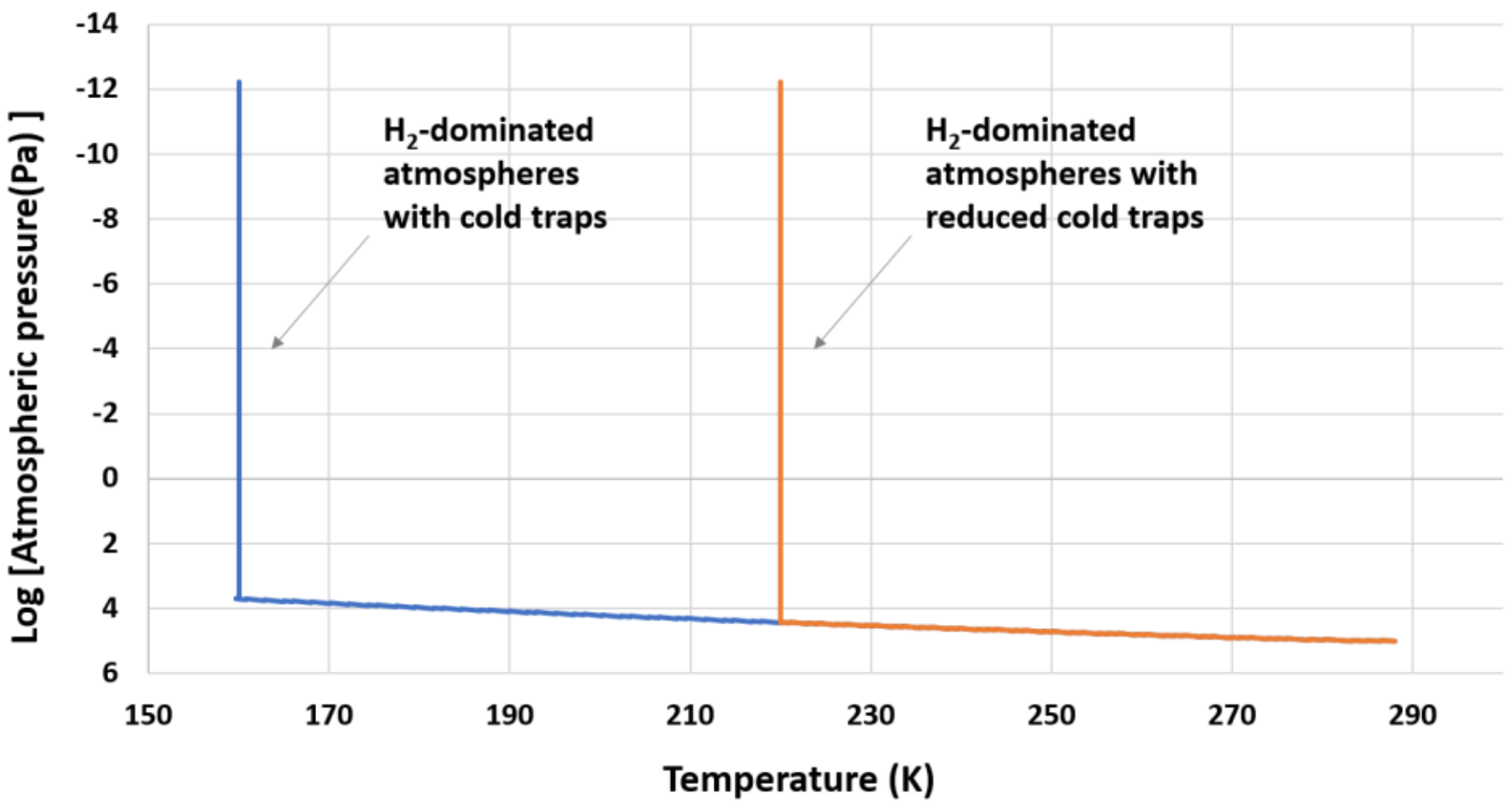

**Figure SI.7**. The temperature-pressure profiles of the simulated exoplanets with 'standard' and 'reduced' cold traps. The y-axis shows atmospheric pressure (Pa) on a log scale, and the x-axis shows temperature in Kelvin (K).

Our results show that when a cold trap is reduced, the photochemical production of $NH_3$ from $NH_2$ becomes more efficient due to more H radicals being present in the $H_2$-dominated atmosphere. As a result, with the same surface production flux, $NH_3$ column-averaged mixing ratio increases slightly from 5.0 ppm to 5.1 ppm (see Table S6-2).

**Table S6-2.** Simulated mixing ratios and reaction rates for exoplanets with $H_2$-dominated atmospheres orbiting M dwarf stars (M5V).

| | With a cold trap | Without a cold trap |
|---|---|---|
| $NH_3$ outgassing [molecule/(cm$^2$s)] | $6.40 \times 10^{15}$ | $6.40 \times 10^{15}$ |
| $NH_3$ column-averaged mixing ratio | 5.0 ppm | 5.1 ppm |
| Chemical production [molecule/(cm$^2$s)] | $4.90 \times 10^{9}$ | $1.20 \times 10^{10}$ |
| Chemical loss [molecule/(cm$^2$s)] | $1.54 \times 10^{10}$ | $1.21 \times 10^{10}$ |
| Dry deposition [molecule/(cm$^2$s)] | $4.16 \times 10^{14}$ | $4.16 \times 10^{14}$ |
| Wet deposition [molecule/(cm$^2$s)] | $5.98 \times 10^{15}$ | $5.98 \times 10^{15}$ |

We currently include a cold trap in each of our existing photochemical simulations. To simulate cold traps, we set the temperature above tropopause to 160 K, 200 K, and 175 K for the $H_2$-dominated, $N_2$-dominated, and $CO_2$-dominated atmosphere. In comparison, the Earth's tropopause temperature is about 217 K (US Standard Atmosphere 1976). Please note that a cold trap does not require a thermal inversion. A cold trap is a part of the (upper) atmosphere where the temperature is low enough to condense volatiles like water (Wordsworth et al., 2013, Wordsworth et al., 2014).

II. The choice of eddy diffusion coefficient ($K_{zz}$): The choice of $K_{zz}$ can affect $NH_3$ concentration because photochemistry is most important in the upper atmosphere layers, whereas $NH_3$ is primarily produced at the planet's surface. $K_{zz}$ is measured or inferred for solar system planets but is not known for exoplanets. The $K_{zz}$ variation does not change our main conclusion, as explained below.

The $K_{zz}$ as a function of altitude (i.e., the eddy diffusion profile) is an input to our code. Each simulated atmospheric scenario has its own fixed eddy diffusion profile. Here, we perform a sensitivity test to a changing $K_{zz}$ for the $H_2$-dominated atmosphere scenario (see Table S6-3).

**Table S6-3.** Steady-state simulation outputs as a function of eddy diffusion magnitude, $NH_3$ surface flux, and presence/absence of wet and dry deposition of $NH_3$ for exoplanets with $H_2$-dominated atmospheres orbiting M dwarf stars (M5V).

| With $NH_3$ deposition | | | |
|---|---|---|---|
| Eddy diffusion coefficient scaling factor | 0.1 | 1 | 10 |
| $NH_3$ outgassing [molecule/(cm$^2$s)] | $6.40 \times 10^{15}$ | $6.40 \times 10^{15}$ | $6.40 \times 10^{15}$ |
| $NH_3$ column-averaged mixing ratio | 3.3 ppm | 5.0 ppm | 12 ppm |
| Chemical production [molecule/(cm$^2$s)] | $1.78 \times 10^{9}$ | $4.90 \times 10^{9}$ | $6.05 \times 10^{9}$ |
| Chemical loss [molecule/(cm$^2$s)] | $1.79 \times 10^{9}$ | $1.54 \times 10^{10}$ | $2.44 \times 10^{10}$ |
| Dry deposition [molecule/(cm$^2$s)] | $4.26 \times 10^{14}$ | $4.16 \times 10^{14}$ | $3.93 \times 10^{14}$ |
| Wet deposition [molecule/(cm$^2$s)] | $5.97 \times 10^{15}$ | $5.98 \times 10^{15}$ | $6.01 \times 10^{15}$ |
| Without $NH_3$ deposition | | | |
| Eddy diffusion coefficient scaling factor | 0.1 | 1 | 10 |
| $NH_3$ outgassing [molecule/(cm$^2$s)] | $1.44 \times 10^{10}$ | $1.44 \times 10^{10}$ | $1.44 \times 10^{10}$ |
| $NH_3$ column-averaged mixing ratio | 9.8 ppm | 5.0 ppm | 1.9 ppm |
| Chemical production [molecule/(cm$^2$s)] | $5.76 \times 10^9$ | $5.52 \times 10^9$ | $6.37 \times 10^9$ |
| Chemical loss [molecule/(cm$^2$s)] | $2.02 \times 10^{10}$ | $2.00 \times 10^{10}$ | $2.08 \times 10^{10}$ |
| Dry deposition [molecule/(cm$^2$s)] | 0 | 0 | 0 |
| Wet deposition [molecule/(cm$^2$s)] | 0 | 0 | 0 |

We find that the $K_{zz}$ variation does not affect rainout or dry deposition. In a more diffusive atmosphere (i.e., larger $K_{zz}$), the photochemical recycling of $NH_3$ becomes more efficient. Specifically, as $K_{zz}$ increases, the photochemical loss of $NH_3$ becomes larger, mainly due to enhanced $NH_3$ direct photolysis ($NH_3 \rightarrow NH_2 + H$). Simultaneously, the reproduction of $NH_3$ from $NH_2$ also increases ($H_2 + NH_2 \rightarrow NH_3 + H$) since there are more $NH_2$ radicals in the atmosphere. Our findings are consistent with (Kasting et al., 1982).

The effect of $K_{zz}$ variation on $NH_3$ atmosphere abundance depends on whether there are $NH_3$-removal sinks on the surface. When $NH_3$ deposition is present (i.e., the surface is not saturated with $NH_3$), the column-averaged mixing ratio of $NH_3$ scales with $K_{zz}$. Compared to chemical loss, wet and dry deposition can remove $NH_3$ much more efficiently than chemical loss (see Table S6-3). As $K_{zz}$ increases, more $NH_3$ is transported to the upper atmosphere before $NH_3$ can be removed by deposition. In this case, larger $K_{zz}$ suppresses the efficacy of $NH_3$ deposition, leading to an increase in atmospheric $NH_3$ concentration. However, when the surface is saturated with $NH_3$ (i.e., without $NH_3$ deposition), $NH_3$ column-averaged mixing ratio is inversely proportional to $K_{zz}$. In a more stagnant atmosphere (i.e., smaller $K_{zz}$), more $NH_3$ can accumulate at low altitudes since there is no wet or dry deposition. This local accumulation of $NH_3$ at low altitudes causes $NH_3$ surface mixing ratio and $NH_3$ column-averaged mixing ratio to increase. As $K_{zz}$ increases, the $NH_3$ that could have accumulated locally at low altitudes is now transported to the upper atmosphere, where $NH_3$ photochemical removal dominates. Hence, when there is no deposition, $NH_3$ column-averaged mixing ratio decreases when the atmosphere becomes more diffusive.

Overall, the $K_{zz}$ variation does not change our main conclusion as $NH_3$ column-averaged mixing ratio stays roughly at the same order of magnitude in our sensitivity test.

III. Henry's law constant for $NH_3$: The effect of pH on $NH_3$ Henry's law constant is minimal. Even under high pH conditions (pH~14), $NH_3$ Henry's law constant will only decrease by up to 10% (Shi et al., 1999).

What might conceivably be quite different is the effective Henry's law constant invoked in the rainout calculation, which assumes a raindrop pH of 5 (Giorgi et al., 1985; Hu et al., 2012). At such acidic pH, the effective Henry's law constant (including partitioning of N(-III) into $NH_4^+$) is much higher than Henry's law constant alone. As raindrop becomes basic (pH>7), the effective Henry's law constant will converge with Henry's law constant. Hence, rainout would be much less efficient (though not ineffective, as even the 'regular' Henry's law constant for $NH_3$ is very high). In this case, we expect results intermediate to our $NH_3$ deposition on and off cases.

## SI.7 $NH_3$ Condensation in the Atmosphere

We find that $NH_3$ does not condense in our simulated atmospheres. At even the highest $NH_3$ flux considered in our study, $NH_3$ is insufficiently abundant anywhere in the atmosphere to condense. Here we plot the scenario with the highest $NH_3$ surface flux (i.e., $H_2$-dominated atmosphere, with $NH_3$ deposition), and we superimpose the $NH_3$ saturation curve (equilibrium with both the liquid and solid phase) on top of it (see Figure SI.8). The black curve is the $NH_3$ volume mixing ratio on planets with $H_2$-dominated atmospheres orbiting M dwarfs. The red curve is the $NH_3$ saturation curve (Stull, 1947). The $NH_3$ surface production flux is $6.4\times10^{15}$ molecule $cm^{-2}$ $s^{-1}$ (see Table 4-1).

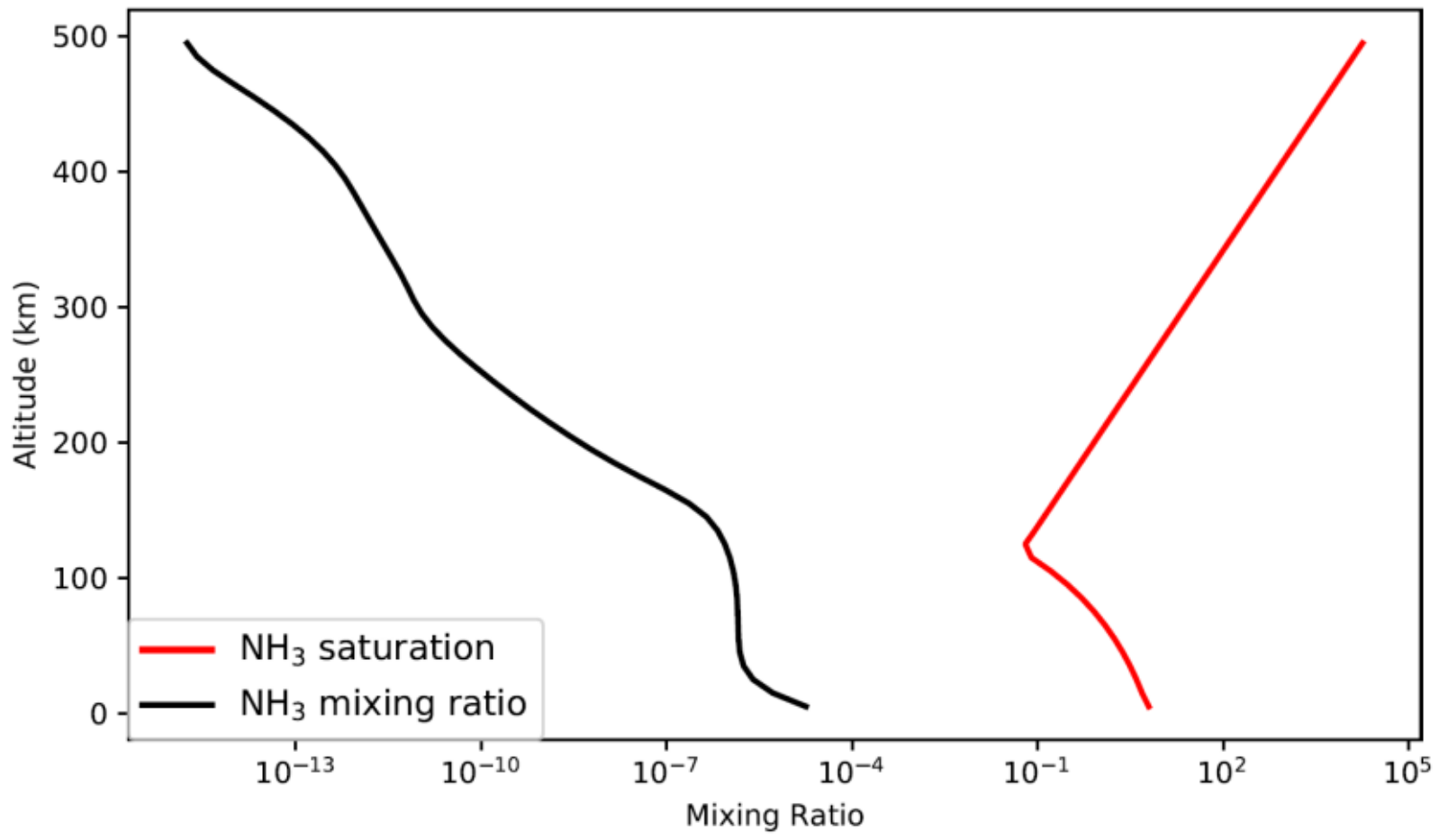

**Figure SI.8**. The volume mixing ratio of $NH_3$ on planets with $H_2$-dominated atmospheres orbiting active M dwarfs. The $NH_3$ surface production flux is $6.4\times10^{15}$ molecules $cm^{-2}$ $s^{-1}$. The y-axis shows altitude in km, and the x-axis shows the mixing ratio. At even the highest $NH_3$ flux considered in our study, $NH_3$ is insufficiently abundant anywhere in the atmosphere to condense.

This figure clearly shows that our simulated $NH_3$ is not concentrated enough to condense in the atmosphere. Therefore, the stratospheric mixing ratio of $NH_3$ will never decouple from the surface mixing ratio (or the surface production flux). As a result, we do not consider upper atmosphere $NH_3$ condensation in our study.

## SI.8 $NH_3$ Abiotic False-Positive Analysis

I. Volcanic $NH_3$ outgassing: We can estimate the flux at which $NH_3$ can be degassed on an exoplanet based on Earth's $N_2$ volcanic outgassing flux. (Catling et al., 2017) estimates the $N_2$ volcanic outgassing on Earth is roughly $0.9\times10^{11}$ mol/year. If all the nitrogen comes out as $NH_3$ instead, the $NH_3$ outgassing flux can be as high as $6.7\times10^{8}$ molecules $cm^{-2}$ $s^{-1}$. For an exoplanet with $H_2$-dominated atmospheres, to generate an $NH_3$ false positive from volcanic outgassing, the exoplanet's mantle conditions need to favor $NH_3/NH_4^+$ production, and the volcanic outgassing of N needs to be at least 100 times greater than Earth's (assuming there is no $NH_3$-removal sink on the planet's surface). Hence, it is unlikely for volcanic $NH_3$ outgassing alone to generate a detectable false positive.

II. Nitrogen photoreduction on $TiO_2$ containing sands: We can estimate the upper limit of the amount of $NH_3$ this abiotic $N_2$ photo-fixation can produce. Suppose there is an Earth-sized exoplanet orbiting an M dwarf. The planet is arid, and the surface is covered by $TiO_2$ containing sands. We assume the planet has an Earth-sized desert (about $1.9\times10^7$ square miles) (Schrauzer et al., 1983). We also assume the planet's total surface area is the same as Earth's (roughly $1.97\times10^8$ square miles). Since M dwarf planets receive roughly 100 ~1000 times less UV than Sun-like star planets do (Ranjan et al., 2017), we assume this abiotic pathway is only 1% effective on the M dwarf planets. Hence, this abiotic $N_2$ fixation can produce up to 1 Tg of $NH_3$ every year (about $2.0\times10^8$ molecules $cm^{-2}$ $s^{-1}$), roughly 2 orders of magnitude smaller than the flux needed to generate a detectable $NH_3$ false positive (assuming there is no $NH_3$-removal sink on the surface). (Kasting et al., 1982) reports that this abiotic $N_2$ photo-fixation can generate $NH_3$ at a rate of $2.8\times10^{10}$ to $2.8\times10^{11}$ molecules $cm^{-2}$ $s^{-1}$ locally on Earth. Assuming this reaction is 1% effective on the M dwarf planets, we estimate that this abiotic $N_2$ fixation could produce $NH_3$ at a rate of up to $2.8\times10^9$ molecules $cm^{-2}$ $s^{-1}$. Even in the most optimum case where $NH_3$ is emitted globally at this rate on exoplanets, the $NH_3$ production flux is still about an order of magnitude smaller than the flux needed to generate a detectable false positive. Overall, it is unlikely for nitrogen photoreduction on $TiO_2$ containing sands alone to generate a detectable false positive.

III. Abiotic $NH_3/NH_4^+$ synthesis around Hadean submarine hydrothermal vents: Suppose there is an Earth-sized exoplanet orbiting an M dwarf star. We assume the planet's total heat flow is the same as modern Earth's (roughly $4.3\times10^{13}$ W) (Elderfield et al., 1996; Smirnov et al., 2008). Assuming the most favorable scenario with 80% of the planet's heat flow released via hydrothermal activity and 10% conversion of $N_2$ to $NH_4^+$, the annual $NH_4^+$ production can be as high as $1.9\times10^{12}$ mol/year (Smirnov et al. 2008). If we further assume all the $NH_4^+$ produced in the ocean is released into the atmosphere as $NH_3$, this abiotic $NH_3/NH_4^+$ synthesis can produce up to 32 Tg of $NH_3$ every year (about $7\times10^9$ molecules $cm^{-2}$ $s^{-1}$). Even in this most optimum case, the $NH_3$ production flux is not large enough to generate a detectable $NH_3$ false positive.

## SI.9 Detectability of TRAPPIST-1 Planets with JWST

Recently, some papers (e.g., Fauchez et al., 2019; Lustig-Yaeger et al., 2019) claim that gases like $CH_4$, $H_2O$, $CO_2$, and $SO_2$ might be detectable for TRAPPIST-1 planets with JWST (Lustig-Yaeger et al., 2019). Depends on the assumptions of the atmosphere (i.e., $CO_2$-, $N_2$- or $H_2O$-dominated atmospheres) and the signal-to-noise (S/N) ratio, the number of transits required to detect those gases on TRAPPIST-1 planets could range from 10 transits to more than 100 transits. What they claim is not contradictory to our statement here. TRAPPIST-1, an ultra-cooled M8V dwarf star, is rare and unique. TRAPPIST-1's small stellar radius and relatively low temperature are beneficial for trace gas detection because the transit depth scales with $(R_{planet}/R_{star})^2$. Furthermore, neither (Fauchez et al., 2019) or (Lustig-Yaeger et al., 2019) assumes a fixed systematic noise floor as we do in our analysis. Lastly, the detectability of gases on exoplanets depends on more than a dozen parameters - the presence of clouds and hazes, scattering, atmospheric refraction, to name a few. Different assumptions for those parameters can lead to very different conclusions. It is beyond the scope of this paper to go through every parameter to assess detectability. Our findings are valid given the assessment metrics we mention in Section 3.4.